\documentclass[11pt]{article}

\PassOptionsToPackage{authoryear}{natbib}

\usepackage[authoryear]{natbib}
\usepackage[margin=1.0in]{geometry}

\usepackage{xcolor}
\usepackage[utf8]{inputenc}
\usepackage[T1]{fontenc}
\usepackage[hidelinks]{hyperref}
\usepackage{url}
\usepackage{booktabs}
\usepackage{amsfonts}
\usepackage{nicefrac}
\usepackage{microtype}
\usepackage{enumitem}
\usepackage{amsmath}
\usepackage{bbm}
\usepackage{graphicx}
\usepackage{subcaption}
\usepackage{datetime2}
\usepackage{setspace}
\usepackage{rotating}
\usepackage{comment}
\usepackage{listings}
\usepackage{multirow}
\usepackage{tcolorbox}
\usepackage{tikz}
\usetikzlibrary{arrows.meta, decorations.pathreplacing, calc}

\usepackage{newtxtext}
\usepackage{newtxmath}
\DeclareMathSizes{5}{5}{3.5}{2.5}

\DeclareMathOperator*{\argmax}{arg\,max}

\makeatletter
\let\old@makefntext\@makefntext
\renewcommand{\@makefntext}[1]{\setstretch{1}\old@makefntext{#1}}
\makeatother

\title{\textbf{Rewarding Noise: Incentives \\ in Decentralized A/B Testing}}
\date{}

\author{%
\begin{tabular}{c}
   Ryan Kessler\thanks{Email: \texttt{rykessle@amazon.com}.} \\
   \emph{Amazon}
\end{tabular}
}

\begin{document}

\maketitle

\begin{abstract}
\begin{spacing}{0.95}
  Many firms that run A/B tests struggle with low statistical power, increasing the risk that launch decisions are driven by noise rather than true impact. We argue that in decentralized organizations, low power can reflect misaligned incentives: product teams choose what and how to test on a customer-facing experience to maximize their private reward, while the experience owner sets evaluation rules and launch criteria to optimize the experience. We formalize this interaction via a principal-agent model and show that evaluating tests using raw impact estimates creates a reward to sampling noise. Correctly calibrated posterior-mean evaluation removes this reward and creates a local incentive to reallocate toward fewer, more precise tests of more ambitious ideas. Using data on A/B tests run at Amazon, we document patterns consistent with these model predictions. After an Amazon experience shifted to posterior-mean-based evaluation in 2024, test volume declined while both precision and the dispersion of true impacts increased, more than doubling signal-to-noise ratios. Our results highlight that A/B testing methodologies can shape not only measurement and launch decisions, but which tests are pursued in the first place.
\end{spacing}
\end{abstract}

\clearpage

\section{Introduction}

Many firms that run A/B tests struggle with low statistical power, increasing the risk that launch decisions are driven by noise rather than true impact. When power is low, selection on statistical significance distorts both the sign and magnitude of estimated impacts \citep{gelman_carlin_2014}. This problem is not confined to small or unsophisticated firms. Even those with sophisticated infrastructure running thousands of tests per year report that many tests are underpowered \citep{kohavi_etal_2024}. 

A common explanation for underpowered tests is that experimenters lack proper tooling.\footnote{As \cite{kohavi_etal_2024} note, ``research regarding improving experimental power for [online controlled experiments] tends to focus on boosting sensitivity, either by directly reducing the variance of [outcomes] or by reducing the variance of estimators.''} Many firms have invested heavily in power calculators, sample size recommenders, and variance reduction techniques \citep{kohavi_etal_2024, gupta_etal_2019}. These tools help, with variance reduction techniques alone capable of reducing sampling variances by more than 20 percent \citep{deng_etal_2013, masoero_etal_2023}. Yet underpowered tests persist even where these features are widely available, suggesting that tooling alone cannot explain the problem.

In this paper, we argue that low power can instead reflect misaligned incentives between the teams that run experiments and those that evaluate them. Large-scale A/B testing is typically decentralized: product teams decide what and how to test on a customer-facing experience, while the experience owner --- the team responsible for the end-to-end quality of that experience --- chooses how to evaluate test results (i.e., a ``scoring rule'') and sets criteria for which features can launch. Because teams are rewarded based on the measured impact of their launches rather than their true impact, the two actors' choices interact, with the product team's testing behavior shaping experience owner choices and vice versa. 

We formalize this interaction via a principal-agent model and show how the experience owner's choices shape the product team's testing behavior. We consider two scoring rules. Raw scoring takes estimated impacts at face value. Posterior-mean scoring pulls estimates toward a prior mean in proportion to their sampling noise. The model predicts that raw scoring creates a reward to noise. Under raw scoring, product teams can increase their expected reward by running many low-precision tests, exploiting the chance of a launch with a large estimated impact driven by sampling error. Relative to the product team's optimal testing behavior, the experience owner would prefer a marginal reallocation toward fewer, more precise tests of more ambitious ideas. Correctly calibrated posterior-mean scoring removes the reward to sampling noise and aligns the two actors' objectives when complexity costs associated with launches are zero. Holding the launch threshold fixed, switching to posterior-mean scoring creates an incentive for the product team to reallocate toward fewer, more precise tests of more ambitious ideas. 

We test these model predictions using data on A/B tests run on an Amazon experience before and after it shifted from approximately raw to posterior-mean scoring in 2024. Consistent with model predictions, we estimate that test volume fell, precision improved, and the dispersion of true impacts increased, resulting in a more than two-fold increase in signal-to-noise ratios. Together, these shifts imply an approximately 60 percent increase in the value of testing on the experience, despite more than 50 percent fewer launches. The scoring rule change coincided with a change in launch criteria, so we cannot isolate the effect of the scoring rule per se. However, as a falsification test, we show that there were no such changes on a different Amazon experience, which continued to use raw scoring throughout our analysis period.

Our findings highlight that A/B testing methodologies can shape not only measurement and launch decisions, but which tests are pursued in the first place. For experiences weighing whether to adopt Bayesian methods, for example, the debate typically centers on potential measurement benefits such as lower mean-squared error and reduced winner's curse bias \citep{efron_morris_1971, cator_vanzwet_2021}. We add a consideration beyond measurement: posterior-mean scoring changes the underlying incentives that determine what teams choose to test. As advances in automation and AI reduce the cost of testing, these incentive effects may become more important, as lower costs without appropriate scoring rules can reinforce incentives to pursue noise rather than true impact.

Our work contributes to a growing literature on principal-agent problems in experimentation \citep{chassang_etal_2012, tetenov_2016, min_2023, bates_etal_2024}, which studies how strategic agents respond to statistical protocols, primarily in clinical trials. Closest to our work is \cite{ghosh_taubinsky_2025}, who also study misalignment between product teams and experience owners in the context of A/B testing. In their model, misalignment arises from the product team's reduced-form preference for launching features. We instead ground misalignment in the scoring rule, which experience owners typically control. This allows us to speak directly to practical questions facing experience owners, such as whether to evaluate tests using frequentist or Bayesian methods \citep{deng_2015, schultzberg_franberg_2026}.

Concurrent work by \citet{soumalias_etal_2026} studies a related mechanism-design problem, showing that sample splitting can realign a strategic experimenter's incentives. Their results assume that the experimenter already knows each test's true effect, isolating moral hazard from estimation uncertainty. We instead model ex ante uncertainty over impacts and provide empirical evidence that a real scoring-rule change shifted testing behavior as our model predicts.

\section{Model}

We consider A/B testing in a decentralized organization, where different actors own different parts of the testing process. An experience owner seeks to maximize the performance of a customer-facing experience by choosing how tests are evaluated and which features launch. A product team seeks to maximize its private reward by choosing what and how to test, taking the experience owner's rules as given. The two actors' choices interact, with the product team's testing behavior shaping experience owner choices and vice versa. 

\subsection{Environment}
\label{sec:model_environment}

We formalize the interaction between the experience owner and the product team via a principal-agent model with three stages, building on \cite{ghosh_taubinsky_2025}. In the first stage, the experience owner (``principal'') commits to a scoring rule $s()$ and a launch threshold $\kappa()$. In the second stage, the product team (``agent'') chooses the number of features to test $N$ and the variance of their true impacts $\sigma^2$. In the third stage, uncertainty is realized. True impacts are drawn, tests are evaluated, qualifying features launch, and payoffs are realized.

The product team chooses between incremental and speculative ideas. We parameterize this choice via $\sigma^2$, the variance of the distribution of true impacts. Higher $\sigma^2$ corresponds to ``big bets'' with greater upside potential. Lower $\sigma^2$ corresponds to safer, incremental improvements. We assume that the mean of the true impacts $\mu$ is exogenous to both the product team and the experience owner. This reflects the idea that it may be easier for the product team to influence the variance of true impacts than their mean. Given $(\mu, \sigma^2)$, each feature's true impact is drawn independently from the normal distribution:

\begin{equation}
    \label{eq:prior}
    \Delta \sim N\left(\mu, \sigma^2\right) 
\end{equation}

Once the product team selects which features to test, the experience owner observes the resulting $\sigma^2$. The experience owner's scoring rule and launch threshold can therefore depend on $\sigma^2$. 

Each A/B test delivers an estimate $\hat{\Delta}$ of the true impact $\Delta$. The product team's choice of the number of tests $N$ determines the sampling variance of $\hat{\Delta}$ via a function $\tau^2(N)$ that is strictly increasing in $N$. Holding total testing resources fixed, running more tests reduces the traffic or effort allocated to each test, leading to noisier estimates of impacts.

Motivated by the randomization of the tests and the central limit theorem, we assume that the estimate $\hat{\Delta}$ is normally distributed, with mean $\Delta$ and variance $\tau^2(N)$: 

\begin{equation}
    \label{eq:sampling}
    \hat{\Delta} \big | \Delta, \tau^2(N) \sim N\left(\Delta, \tau^2(N)\right)
\end{equation}

We sometimes suppress the dependence on $N$ and write $\tau^2 = \tau^2(N)$. We let $\phi(\cdot)$ and $\Phi(\cdot)$ denote the standard normal PDF and CDF, respectively. 

The experience owner's scoring rule maps estimated impacts to a score that determines both launch decisions and the product team's reward. We restrict attention to linear shrinkage scoring rules:

\begin{equation}
    \label{eq:scoring}
    s\left(\hat{\Delta}, \tau^2, \sigma^2\right) = \mu + w\left(\tau^2, \sigma^2\right) \cdot \left(\hat{\Delta}  - \mu\right) 
\end{equation}

\noindent where $w(\tau^2, \sigma^2) \geq 0$ is decreasing in $\tau^2$ and increasing in $\sigma^2$. This class nests the two cases central to our analysis. ``Raw scoring'' sets $s(\hat{\Delta}, \tau^2, \sigma^2) = \hat{\Delta}$, while ``posterior-mean scoring'' sets $s(\hat{\Delta}, \tau^2, \sigma^2) =\mu +  \sigma^2(\sigma^2 + \tau^2)^{-1} (\hat{\Delta} - \mu)$.

The experience owner's launch threshold $\kappa()$ determines which features launch on the experience. A feature launches if and only if its score $s(\hat{\Delta}, \tau^2, \sigma^2)$ exceeds $\kappa(\tau^2, \sigma^2)$. For example, under raw scoring $s(\hat{\Delta}, \tau^2, \sigma^2) = \hat{\Delta}$, setting $\kappa(\tau^2, \sigma^2) = \tau \cdot \Phi^{-1}(1- \alpha/2)$ corresponds to launching features if and only if $\hat{\Delta}$ is positive and statistically significant at level $\alpha$. 

\subsection{Product Team's Problem}
\label{sec:model_product_team}

The product team chooses $N$ and $\sigma^2$ to maximize its expected reward from launched features, subject to a resource constraint. Define the launch indicator:

\begin{equation}
    \label{eq:launch_indicator}
    L\left(N, \sigma^2; s, \kappa\right) = 1\left(s\left(\hat{\Delta}, \tau^2(N), \sigma^2\right) > \kappa\left(\tau^2(N), \sigma^2\right)\right)
\end{equation}

The product team's expected reward is:

\begin{equation}
    \Pi_P\left(N, \sigma^2\right) = N \cdot E\left(s\left(\hat{\Delta}, \tau^2(N), \sigma^2\right) \cdot  L\left(N, \sigma^2; s, \kappa\right)\right)
\end{equation}

The product team is rewarded based on the scoring rule $s()$, not the true impact of its launches. No one observes true impacts directly, and teams are instead evaluated using measured test outcomes. 

The product team's problem is:

\begin{equation}
    \label{eq:product_team_problem}
    \left(\tilde{N}(s, \kappa), \tilde{\sigma}^2(s, \kappa)\right) = \argmax_{(N, \sigma^2)} \Pi_P\left(N, \sigma^2\right) \: \: \: \text{subject to} \: \: \: R(N, \sigma^2) \leq \bar{R}
\end{equation}

\noindent where $R(N, \sigma^2)$ is the total resources required to run $N$ tests when the variance of true impacts is $\sigma^2$ and $\bar{R}$ is the product team's total resource budget.

The product team's resource function $R()$ is strictly increasing in both its arguments $N$ and $\sigma^2$. Running additional tests requires greater setup and coordination effort, while pursuing more ambitious ideas requires greater design and development effort. We assume throughout that $\tau^2(\cdot)$ and $R(\cdot)$ are sufficiently well-behaved that both the product team's and the experience owner's problems have interior solutions.

\subsection{Experience Owner's Problem} 
\label{sec:model_experience_owner}

The experience owner chooses $s()$ and $\kappa()$ to maximize the expected true impact of launched features net of complexity costs $c_X$, anticipating the product team's best response $\{\tilde{N}(s, \kappa), \tilde{\sigma}^2(s, \kappa)\}$. 

The experience owner's expected reward is:

\begin{equation}
    \Pi_E\left(s, \kappa\right) = \tilde{N}(s, \kappa) \cdot E\left((\Delta - c_X) \cdot  L\left(\tilde{N}(s, \kappa), \tilde{\sigma}^2(s, \kappa); s, \kappa\right)\right)
\end{equation}

The experience owner's problem is: 

\begin{equation}
    \label{eq:experience_owner_problem}
    \left(\tilde{s}, \tilde{\kappa}\right) = \argmax_{(s(), \kappa())} \Pi_E\left(s, \kappa \right) 
\end{equation}

Unlike the product team, which is rewarded based on the scoring rule $s()$, the experience owner values the true impact of launched features, as they are responsible for the long-term performance of the experience. 

Launched features impose a per-launch complexity cost $c_X \geq 0$ on the experience owner. Each additional feature increases the complexity of the experience through interactions across features and long-run maintenance. The experience owner bears these costs because they are responsible for the experience as a whole. The product team, which is evaluated based on the measured impact of its own launches, does not internalize them.

\section{Theoretical Results} 

The experience owner's choice of scoring rule shapes the product team's testing behavior. We formalize this via four results, holding the launch threshold fixed. We present a proof of each result in Appendix \ref{sec:proofs}.

We begin with raw scoring, where the product team is rewarded based on the estimated impact $\hat{\Delta}$.

\vspace{0.5pc}

\noindent \textbf{Proposition 1 (Raw scoring rewards noise):} Suppose that $\mu \geq 0$ and $\kappa > \mu$. Under equations (\ref{eq:prior}) and (\ref{eq:sampling}) and raw scoring $s(\hat{\Delta}, \tau^2, \sigma^2) = \hat{\Delta}$, at an interior product-team optimum, the experience owner's objective has a strictly positive derivative with respect to $\sigma^2$ along the binding resource frontier. Thus, the experience owner prefers a marginal reallocation toward fewer, more precise tests of more ambitious ideas.

\vspace{0.5pc}

Under raw scoring, the product team is rewarded based on $\hat{\Delta}$, which combines the true impact $\Delta$ and sampling noise. Because the scoring rule does not distinguish between the two, the product team receives credit for favorable sampling noise as well as true impact when a feature meets launch criteria. At the product team's optimum, the experience owner, who cares about true impacts net of complexity costs, can improve its objective through a marginal reallocation toward fewer tests, higher precision, and greater ambition.

\vspace{0.5pc}

\noindent \textbf{Proposition 2 (Vanishing testing costs under raw scoring):} Suppose that $\mu \geq 0$, $\kappa > \mu$, and the resource function takes the form $R(N,\sigma^2)=\gamma N+R_\sigma(\sigma^2)$, where $\gamma>0$. Assume that $R_\sigma$ is continuous and strictly increasing, with $R_\sigma(0)=0$ and $R_\sigma(\bar x)=\bar R$ for some finite $\bar x>0$. Suppose also that $\tau^2(N)\to\infty$ as $N\to\infty$. Under equations (\ref{eq:prior}) and (\ref{eq:sampling}) and raw scoring $s(\hat{\Delta},\tau^2,\sigma^2)=\hat{\Delta}$, any optimal product-team allocation satisfies $\tilde{\sigma}^2\to0$ and $\gamma\tilde N\to\bar R$ as $\gamma\to0$. Thus, the team approaches minimum ambition and maximum feasible test volume, while sampling variance diverges and statistical precision vanishes.

\vspace{0.5pc}

Under these additional assumptions, vanishing testing costs drive the product team toward high-volume, low-precision experimentation. As the product team runs more tests, each test becomes noisier, increasing the chance of a large, noise-induced reward. Raising ambition adds relatively little while costing resources that could instead fund more tests. The team therefore shifts resources from ambition to volume, approaching minimum ambition and maximum feasible test volume. This result illustrates how advances in automation, APIs, and AI could further encourage noise-seeking under raw scoring.

\vspace{0.5pc}

\noindent \textbf{Proposition 3 (Posterior-mean scoring removes the reward to noise):} Suppose that $\mu \geq 0$ and $\kappa > \mu$. Under equations (\ref{eq:prior}) and (\ref{eq:sampling}) and posterior-mean scoring $s(\hat{\Delta}, \tau^2, \sigma^2) = \mu + \sigma^2(\sigma^2 + \tau^2)^{-1}(\hat{\Delta} - \mu)$, the gap $G = \Pi_P - \Pi_E$ between the product team's and the experience owner's objectives is strictly smaller for any $(N, \sigma^2)$ than under raw scoring. When $c_X = 0$, $G = 0$ for all $(N, \sigma^2)$.

\vspace{0.5pc}

Posterior-mean scoring shrinks estimated impacts toward the mean $\mu$, with the degree of shrinkage increasing in the level of sampling noise. As a result, the scoring rule discounts large estimated impacts $\hat{\Delta}$ from low-precision tests while taking the same $\hat{\Delta}$ from a high-precision test at closer to face value. This eliminates the reward to noise identified in Proposition 1. When complexity costs are zero, this aligns the product team's and experience owner's objectives exactly. 

Proposition 1 shows that the product team's raw-scoring allocation is locally inefficient for the experience owner. Proposition 2 shows that, under additional assumptions, vanishing testing costs drive the team toward maximum feasible test volume and minimum ambition. Proposition 3 shows that posterior-mean scoring removes the reward to noise. We now examine how switching scoring rules changes the product team's testing behavior. 

\vspace{0.5pc}

\noindent \textbf{Proposition 4 (Scoring rules change marginal allocation incentives):} Suppose that $\mu \geq 0$ and $\kappa > \mu$. Under equations (\ref{eq:prior}) and (\ref{eq:sampling}), at an interior product-team optimum under raw scoring $s(\hat{\Delta}, \tau^2, \sigma^2) = \hat{\Delta}$, the product team's objective under posterior-mean scoring $s(\hat{\Delta}, \tau^2, \sigma^2) = \mu + \sigma^2(\sigma^2 + \tau^2)^{-1}(\hat{\Delta} - \mu)$ has a strictly positive derivative with respect to $\sigma^2$ along the binding resource frontier. Thus, under posterior-mean scoring, the team prefers a marginal reallocation toward fewer, more precise tests of more ambitious ideas.

\vspace{0.5pc}

Switching from raw to posterior-mean scoring changes the product team's marginal allocation incentives. Starting from an allocation optimal under raw scoring, the team can increase its expected reward under posterior-mean scoring through a marginal reallocation toward fewer tests, higher precision, and greater ambition. This local comparison holds even when $c_X = 0$.

\section{Empirical Application}

Proposition 4 establishes a local incentive to reallocate toward fewer tests, higher statistical precision, and greater ambition when scoring shifts from raw estimates to posterior means. We look for descriptive evidence consistent with this incentive in a change in scoring rules on a high-traffic, customer-facing experience at Amazon (hereafter ``the experience'').

\subsection{Setting and Data}

Multiple product teams run A/B tests on the experience to evaluate new features. The experience owner determines how tests are evaluated and which features launch, with the goal of maximizing a primary metric (``metric A'') while protecting a secondary metric (``metric B''). 

In 2024, the experience owner shifted the experience from approximately raw scoring to posterior-mean scoring. Amazon's central experimentation team, Weblab, has long provided posterior-mean estimates of treatment impacts to experience owners. Prior to 2024, however, these estimates were based on implausibly large values of the variance of true impacts $\sigma^2$, resulting in negligible shrinkage and estimates nearly identical to their raw counterparts \citep{heidemanns_etal_2024}. In February 2024, Weblab updated estimates of $\sigma^2$, enabling more principled posterior-mean scoring \citep{heidemanns_etal_2024}. In May 2024, the experience owner adopted these estimates for scoring and launch decisions and updated their launch criteria. In September 2024, the experience owner began enforcing these posterior-mean-based rules directly in the Weblab application, preventing teams from launching features that did not meet them. 

\begin{figure}[h!]
\centering 
\caption{Scoring rule and launch criteria change: Timeline and analysis periods}
\label{fig:timeline}
\begin{tikzpicture}[
    event/.style={fill=black, circle, inner sep=1.8pt},
    elabel/.style={text width=4.6cm, align=center, font=\footnotesize},
]

\def\SepA{0}       
\def\DecA{1.95}    
\def\FebB{3.25}    
\def\MayB{5.20}    
\def\SepB{7.80}    
\def\DecB{9.75}    
\def\FebC{11.05}   

\fill[blue!10, rounded corners=3pt]
    (\SepA, -0.30) rectangle (\FebB, 0.30);
\fill[green!14, rounded corners=3pt]
    (\SepB, -0.30) rectangle (\FebC, 0.30);

\draw[-{Stealth[length=5pt]}, thick, black!50]
    (-0.8, 0) -- (12.0, 0);

\foreach \x in {\SepA,\DecA,\FebB,\MayB,\SepB,\DecB,\FebC}{
    \draw (\x, -0.08) -- (\x, 0.08);
}

\node[below=5pt, font=\scriptsize] at (\SepA,0) {Sep 2023};
\node[below=5pt, font=\scriptsize] at (\DecA,0) {Dec 2023};
\node[below=5pt, font=\scriptsize] at (\FebB,0) {Feb 2024};
\node[below=5pt, font=\scriptsize] at (\MayB,0) {May 2024};
\node[below=5pt, font=\scriptsize] at (\SepB,0) {Sep 2024};
\node[below=5pt, font=\scriptsize] at (\DecB,0) {Dec 2024};
\node[below=5pt, font=\scriptsize] at (\FebC,0) {Feb 2025};

\draw[decorate,
      decoration={brace, amplitude=5pt, mirror},
      thick, blue!60!black]
    (\SepA, -0.72) -- (\FebB, -0.72)
    node[midway, below=7pt,
         font=\footnotesize \bfseries, blue!60!black]
    {Pre-change period};

\draw[decorate,
      decoration={brace, amplitude=5pt, mirror},
      thick, green!60!black]
    (\SepB, -0.72) -- (\FebC, -0.72)
    node[midway, below=7pt,
         font=\footnotesize \bfseries, green!60!black]
    {Post-change period};

\draw[{Stealth[length=4pt]}-{Stealth[length=4pt]},
      thick, dashed, black!35]
    ({(\SepA+\FebB)/2}, -1.65) -- ({(\SepB+\FebC)/2}, -1.65)
    node[midway, below=2pt,
         font=\scriptsize, black!70]
    {Year-over-year comparison to account for potential seasonality};

\node[event] (e0) at (\DecA, 0) {};
\node[event] (e1) at (\FebB, 0) {};
\node[event] (e2) at (\MayB, 0) {};
\node[event] (e3) at (\SepB, 0) {};

\node[elabel] (l0) at (\DecA, 2.7)
    {\tiny Weblab announces plan to \\ [-3.5pt]
     launch new Bayesian priors};
\draw[thin, black!35] (e0.north) -- (l0.south);

\node[elabel] (l1) at (\FebB, 1.65)
    {\tiny Weblab launches new \\ [-3.5pt]
     Bayesian priors};
\draw[thin, black!35] (e1.north) -- (l1.south);

\node[elabel] (l2) at (\MayB, 2.7)
    {\tiny Experience owner announces \\  [-3.5pt]
     new launch criteria};
\draw[thin, black!35] (e2.north) -- (l2.south);

\node[elabel] (l3) at (\SepB, 1.65)
    {\tiny Experience owner enforces new \\ [-3.5pt]
     launch criteria via Weblab App};
\draw[thin, black!35] (e3.north) -- (l3.south);
\end{tikzpicture}
\end{figure}
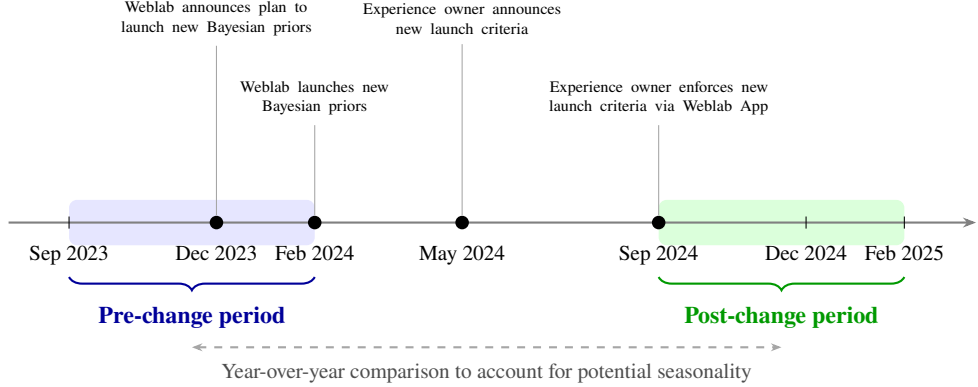

Figure \ref{fig:timeline} presents the timeline of these changes as well as the time periods we use in our analysis. We compare testing behavior in two five-month windows: September 2023 through January 2024 (pre-change period) and September 2024 through January 2025 (post-change period). The pre-change period ends in January 2024, before Weblab launched its updated estimates of $\sigma^2$. The post-change period begins in September 2024, when the experience owner began enforcing its posterior-mean-based rules directly in the Weblab application. The year-over-year comparison accounts for potential seasonality in testing behavior. Within each period, we restrict attention to tests run in the US marketplace whose most recent start date falls within the period. For each test, we restrict attention to estimated impacts on metrics A and B as of the longest test duration. 

\subsection{Estimation}

For both the pre- and post-change periods, we estimate the model's three parameters as follows. We estimate the number of tests $N$ as the number of distinct test-treatment pairs. We estimate the sampling standard error $\tau$ as the median of the estimated standard errors of the estimated difference in means across test-treatment pairs. We estimate the variance of true impacts $\sigma^2$ via maximum likelihood under the normality assumptions in equations (\ref{eq:prior}) and (\ref{eq:sampling}). We report 95 percent confidence intervals on the post-to-pre ratio of our estimates. For $\tau$, we use a nonparametric bootstrap, while for $\sigma$ and $\sigma/\tau$ we rely on standard asymptotic inference. 

\subsection{Results}

\begin{figure}[h!]
    \caption{Changes in testing behavior following the scoring rule and launch criteria change}
    \label{fig:empirical_results}
    \centering 
    \begin{tabular}{ccc}
    (a) Metric A & (b) Metric B  \\
    \includegraphics[scale = 0.45]{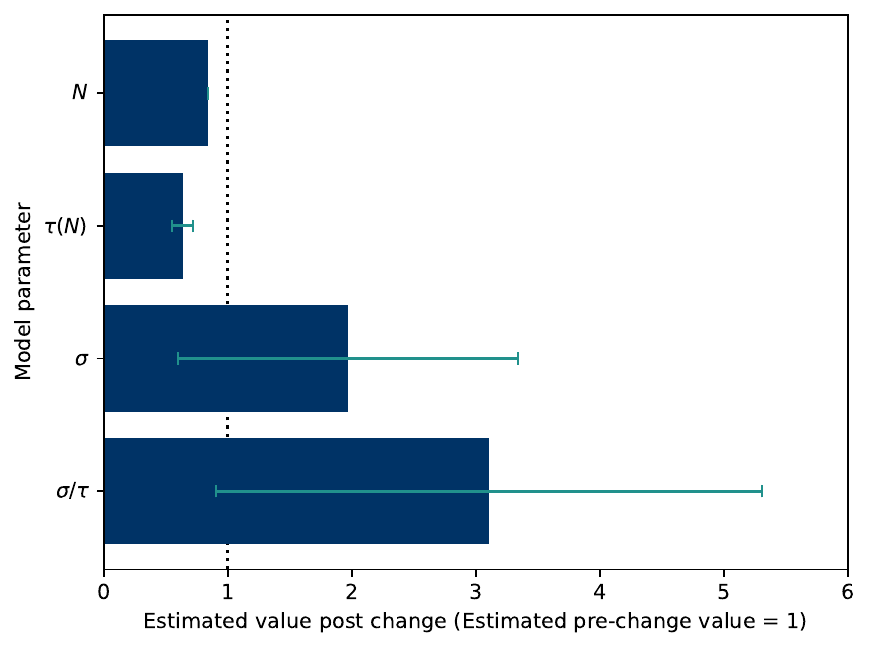} & \includegraphics[scale = 0.45]{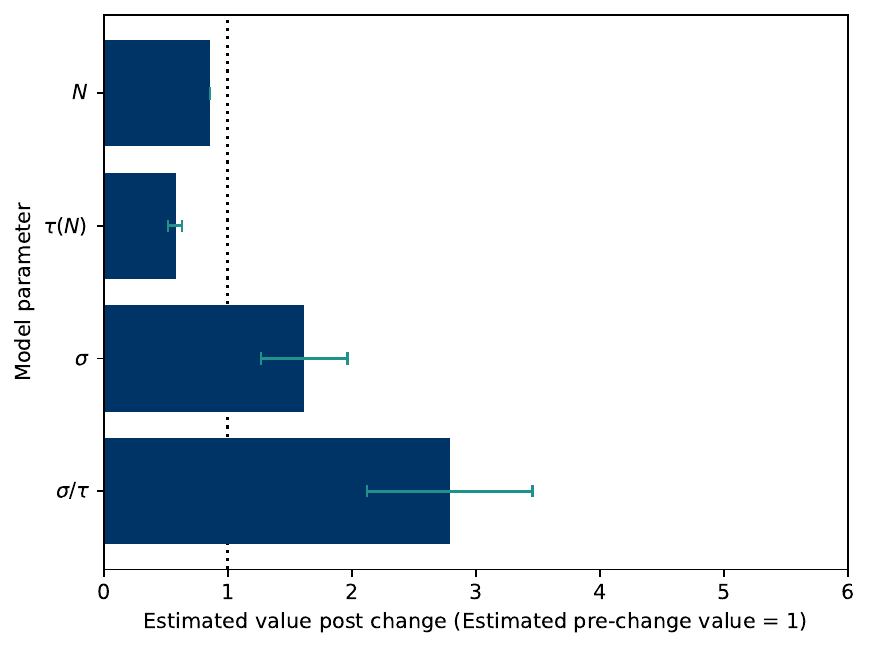}
    \end{tabular}
    \begin{flushleft}
        \tiny
        \emph{Notes:} Each bar reports the ratio of the post-change to pre-change estimate of the indicated parameter. The dotted line marks a ratio of one (no change). Error bars denote 95 percent confidence intervals.
    \end{flushleft}
\end{figure}

Figure \ref{fig:empirical_results} presents evidence consistent with the model's predictions. Panel (a) presents results for metric A, while panel (b) presents results for metric B. For both metrics, the number of tests $N$ fell by approximately 15 percent and the sampling standard error $\tau$ fell by over 30 percent. The estimated standard deviation of true impacts $\sigma$ roughly doubled for metric A and increased by around 60 percent for metric B, though the metric A estimate is imprecise due to a low signal-to-noise ratio in the pre-change period. Together, these shifts imply a more than two-fold increase in the signal-to-noise ratio $\sigma/\tau$ for both metrics, with the metric B estimate statistically significant.

These findings do not appear to be driven by outliers, compositional shifts, or secular trends. Figure \ref{fig:robustness} shows that the results are stable across outlier trimming, compositional reweighting, exclusion of the largest product teams, alternative time windows, and alternative outcome metrics. As a falsification test, we repeat our analysis using tests run on a different Amazon experience that continued to use raw scoring throughout our analysis period. On this experience, test volume \emph{increased}, while precision and the dispersion of true impacts --- and therefore the signal-to-noise ratio --- went largely unchanged.

These parameter shifts, together with the change in launch criteria, imply a substantial increase in the value of testing on the experience. Choosing $\kappa$ in each period to approximate observed launch rates and setting $c_X = 0$, the value of testing increased by approximately 60 percent for both metrics, despite more than 50 percent fewer launches.

\section{Limitations}
\label{sec:limitations}

Our model makes several simplifying assumptions. First, for tractability, we assume that the true impacts are normally distributed. In practice, the distribution of true impacts may have fatter-than-normal tails \citep{azevedo_etal_2020}. If this is the case, the noise premium under raw scoring may be smaller or larger, and posterior-mean scoring may be more or less effective in mitigating it. Second, we model the ambition of ideas via a single parameter $\sigma^2$. In practice, product teams choose across ideas that differ in many ways, not just their variance. Third, in establishing our main results, we hold the launch threshold $\kappa$ fixed and vary the scoring rule. If the experience owner jointly optimizes both instruments, the optimal launch threshold could partially compensate for a suboptimal scoring rule, reducing the distortion relative to what our model predicts.

Our empirical analysis also has important limitations. First, the change in the scoring rule for tests run on the experience coincided with a change in launch criteria, making it difficult to isolate the effect of the scoring rule change per se. Second, even if the scoring rule change were the only change, estimating its causal effect would require ruling out other concurrent changes --- such as reorganizations --- that could independently drive changes in equilibrium testing behavior. The falsification test on the experience that did not change its scoring rule helps mitigate this concern but does not eliminate it. Third, our estimates of some model parameters are imprecise. The estimate of $\sigma$ is particularly imprecise in the pre-change period for some metrics, where the low signal-to-noise ratio makes it difficult to distinguish the variance of true impacts from zero. 

\bibliographystyle{aer}
\bibliography{bibliography.bib}

\clearpage

\appendix 
\renewcommand{\thefigure}{\thesection\arabic{figure}}
\setcounter{figure}{0}

\clearpage 

\section{Appendix: Proof of Claims}
\label{sec:proofs}

\noindent \textbf{Lemma 1 (Product team's objective under normality):} Under equations (\ref{eq:prior}) and (\ref{eq:sampling}), for any scoring rule $s()$ satisfying equation (\ref{eq:scoring}) and launch threshold $\kappa() > 0$, evaluated at an allocation with $w(N,\sigma^2)>0$, the product team's objective reduces to:

\vspace{-0.25pc}
{\footnotesize
\begin{equation}
    \Pi_P\left(N, \sigma^2\right) =  N \left( \mu \cdot\Phi\left(\frac{\mu - \kappa(N, \sigma^2)}{w\left(N, \sigma^2\right) \nu\left(N, \sigma^2\right)} \right) + w(N, \sigma^2) \nu(N, \sigma^2) \cdot \phi\left(\frac{\kappa(N, \sigma^2) - \mu}{w\left(N, \sigma^2\right) \nu\left(N, \sigma^2\right)}\right) \right)  
\end{equation}}

\noindent where $\kappa(N, \sigma^2) = \kappa(\tau^2(N), \sigma^2)$, $w(N, \sigma^2) = w(\tau^2(N), \sigma^2)$, and $\nu^2(N, \sigma^2) = \sigma^2 + \tau^2(N)$.

\noindent \emph{Proof:} To simplify notation, define $S = s\left(\hat{\Delta}, \tau^2(N), \sigma^2\right)$. By the definition of the conditional expectation:

\vspace{-0.75pc}
\begin{align}
    \Pi_P\left(N, \sigma^2\right) &= N \cdot E\left(S \cdot  L\left(N, \sigma^2; s, \kappa\right)\right) \\
    &= N \cdot \text{Pr}\left(S > \kappa\left(N, \sigma^2\right) \right) E\left(S | S >  \kappa\left(N, \sigma^2\right) \right) \label{eq:lemma1_1}
\end{align}

\noindent Under equations (\ref{eq:prior}) and (\ref{eq:sampling}), for any scoring rule $s()$ satisfying equation (\ref{eq:scoring}), we have that:

\begin{equation}
    S \sim N\left(\mu, w^2(N, \sigma^2) \nu^2(N, \sigma^2) \right) 
\end{equation}

\noindent Therefore, we have that:

\vspace{-0.75pc}
\begin{align}
     \text{Pr}\left(S > \kappa\left(N, \sigma^2\right) \right)  &= 1 - \text{Pr}\left(S  \leq \kappa\left(N, \sigma^2\right) \right) \\
    &= 1 - \text{Pr}\left( \frac{S - \mu}{w(N, \sigma^2)\nu(N, \sigma^2)} \leq \frac{\kappa\left(N, \sigma^2\right) - \mu}{w(N, \sigma^2)\nu(N, \sigma^2)} \right) \\
    &= 1 - \Phi\left(\frac{\kappa\left(N, \sigma^2\right) - \mu}{w\left(N, \sigma^2\right) \nu\left(N, \sigma^2\right)} \right) \label{eq:lemma1_2}
\end{align}

Recalling the truncated-normal property that for any random $X \sim N(\theta, \zeta^2)$ and constant $c \in \mathbb{R}$: 

\begin{equation}
    E(X | X > c) = \theta + \zeta \left(\frac{\phi\left(\frac{c - \theta}{\zeta}\right)}{1 - \Phi\left(\frac{c - \theta}{\zeta}\right)}\right) \label{eq:lemma_1_trunc_normal}
\end{equation}

we have that: 

\begin{equation}
    E\left(S \Big | S > \kappa\left(N, \sigma^2\right) \right) = \mu + w\left(N, \sigma^2\right) \nu\left(N, \sigma^2\right) \left(\frac{\phi\left(\frac{\kappa\left(N, \sigma^2\right) - \mu}{w\left(N, \sigma^2\right) \nu\left(N, \sigma^2\right)} \right)}{1 - \Phi\left(\frac{\kappa\left(N, \sigma^2\right) - \mu}{w\left(N, \sigma^2\right) \nu\left(N, \sigma^2\right)} \right) }\right) \label{eq:lemma1_3}
\end{equation}

Plugging equations (\ref{eq:lemma1_2}) and (\ref{eq:lemma1_3}) into equation (\ref{eq:lemma1_1}) while recognizing that for the normal CDF $1 - \Phi(x) = \Phi(-x)$ for any $x \in \mathbb{R}$, establishes the claim $\hfill \square$

\vspace{0.5pc}

\noindent \textbf{Lemma 2 (Experience owner's objective under normality):} Under equations (\ref{eq:prior}) and (\ref{eq:sampling}), for any scoring rule $s()$ satisfying equation (\ref{eq:scoring}) and launch threshold $\kappa() > 0$, evaluated at an allocation with $w(N,\sigma^2)>0$, the experience owner's objective reduces to:

\vspace{-0.25pc}
{\footnotesize
\begin{equation}
    \Pi_E\left(s, \kappa\right) =  N \left( \left( \mu - c_X\right) \cdot\Phi\left(\frac{\mu - \kappa(N, \sigma^2)}{w\left(N, \sigma^2\right) \nu\left(N, \sigma^2\right)} \right) + \frac{\sigma^2}{\nu(N, \sigma^2)} \cdot \phi\left(\frac{\kappa(N, \sigma^2) - \mu}{w\left(N, \sigma^2\right) \nu\left(N, \sigma^2\right)}\right) \right)  
\end{equation}}

\noindent where $\kappa(N, \sigma^2) = \kappa(\tau^2(N), \sigma^2)$, $w(N, \sigma^2) = w(\tau^2(N), \sigma^2)$, and $\nu^2(N, \sigma^2) = \sigma^2 + \tau^2(N)$.

\noindent \emph{Proof:} To simplify notation, define $S = s\left(\hat{\Delta}, \tau^2(N), \sigma^2\right)$. By the definition of the conditional expectation:

\vspace{-0.75pc}
\begin{align}
    \Pi_E\left(s, \kappa\right) &= N \cdot E\left(\left(\Delta - c_X\right) \cdot  L\left(N, \sigma^2; s, \kappa\right)\right) \\
    &= N \cdot \text{Pr}\left(S > \kappa\left(N, \sigma^2\right) \right) E\left(\Delta - c_X | S >  \kappa\left(N, \sigma^2\right) \right) \label{eq:lemma2_1}
\end{align}

\noindent Under equations (\ref{eq:prior}) and (\ref{eq:sampling}), for any scoring rule $s()$ satisfying equation (\ref{eq:scoring}), we have that:

\begin{equation}
    S \sim N\left(\mu, w^2(N, \sigma^2) \nu^2(N, \sigma^2) \right) 
\end{equation}

\noindent Therefore, we have that:

\vspace{-0.75pc}
\begin{align}
     \text{Pr}\left(S > \kappa\left(N, \sigma^2\right) \right)  &= 1 - \text{Pr}\left(S  \leq \kappa\left(N, \sigma^2\right) \right) \\
    &= 1 - \text{Pr}\left( \frac{S - \mu}{w(N, \sigma^2)\nu(N, \sigma^2)} \leq \frac{\kappa\left(N, \sigma^2\right) - \mu}{w(N, \sigma^2)\nu(N, \sigma^2)} \right) \\
    &= 1 - \Phi\left(\frac{\kappa\left(N, \sigma^2\right) - \mu}{w\left(N, \sigma^2\right) \nu\left(N, \sigma^2\right)} \right) \label{eq:lemma2_2}
\end{align}

Next, note that $(\Delta,S)$ is jointly normal. Under equations (\ref{eq:prior}) and (\ref{eq:sampling}), we have that $\hat{\Delta} = \Delta + \varepsilon$, with $\Delta \sim N(\mu,\sigma^2)$, $\varepsilon \sim N(0,\tau^2(N))$, and $\Delta \perp \varepsilon$. Therefore, $(\Delta,\varepsilon)$ is jointly normal. Since, under equation (\ref{eq:scoring}) the score $S$ is a linear function of $\hat{\Delta}$, it follows that $(\Delta,S)$ is also jointly normal.

By the linear conditional expectation formula for jointly normal random variables, we have that: 

\vspace{-0.75pc}
\begin{align}
E\left(\Delta \mid S\right) &= \mu + \frac{\mathrm{Cov}(\Delta,S)}{\mathrm{Var}(S)}(S-\mu) \\
&=\mu + \frac{\sigma^2}{w\left(N,\sigma^2\right)\,\nu^2\left(N,\sigma^2\right)}(S-\mu).
\end{align}

By the law of iterated expectations: 

\vspace{-0.75pc}
\begin{align}
E\left(\Delta - c_X \mid S > \kappa\left(N,\sigma^2\right)\right) &= E\left(E\left(\Delta - c_X \mid S\right) \mid S > \kappa\left(N,\sigma^2\right)\right)\\ 
& = \mu - c_X + \frac{\sigma^2}{w\left(N,\sigma^2\right)\,\nu^2\left(N,\sigma^2\right)} E\left(S-\mu \mid S > \kappa\left(N,\sigma^2\right)\right)
\end{align}

Recalling the truncated normal property in equation (\ref{eq:lemma_1_trunc_normal}):

\begin{equation}
E\left(S-\mu \mid S > \kappa\left(N,\sigma^2\right)\right) = w\left(N,\sigma^2\right)\,\nu\left(N,\sigma^2\right) \frac{\phi\!\left(\frac{\kappa\left(N,\sigma^2\right)-\mu}{w\left(N,\sigma^2\right)\,\nu\left(N,\sigma^2\right)}\right)}{1-\Phi\!\left(\frac{\kappa\left(N,\sigma^2\right)-\mu}{w\left(N,\sigma^2\right)\,\nu\left(N,\sigma^2\right)}\right)}
\end{equation}

Substituting yields:

\begin{equation}
E\left(\Delta - c_X \mid S > \kappa(N,\sigma^2)\right) = \mu - c_X + \frac{\sigma^2}{\nu(N,\sigma^2)}\frac{\phi\!\left(\frac{\kappa(N,\sigma^2)-\mu}{w(N,\sigma^2)\,\nu(N,\sigma^2)}\right)}{1-\Phi\!\left(\frac{\kappa(N,\sigma^2)-\mu}{w(N,\sigma^2)\,\nu(N,\sigma^2)}\right)} \label{eq:lemma2_3}
\end{equation}

Plugging equations (\ref{eq:lemma2_2}) and (\ref{eq:lemma2_3}) into equation (\ref{eq:lemma2_1}) while recognizing that for the normal CDF $1 - \Phi(x) = \Phi(-x)$ for any $x \in \mathbb{R}$, establishes the claim $\hfill \square$

\vspace{0.5pc}

\noindent \textbf{Proposition 1 (Raw scoring rewards noise):} Suppose that $\mu \geq 0$ and $\kappa > \mu$. Under equations (\ref{eq:prior}) and (\ref{eq:sampling}) and raw scoring $s(\hat{\Delta}, \tau^2, \sigma^2) = \hat{\Delta}$, at an interior product-team optimum, the experience owner's objective has a strictly positive derivative with respect to $\sigma^2$ along the binding resource frontier. Thus, the experience owner prefers a marginal reallocation toward fewer, more precise tests of more ambitious ideas.

\noindent \emph{Proof:} Fix the launch threshold $\kappa$ and assume that $\mu \geq 0$ and $\kappa  > \mu$. By Lemmas 1 and 2, we have that under raw scoring:

\vspace{-0.75pc}
\begin{align}
    \Pi_P &= N\left(\mu \cdot\Phi(-z)+\nu \cdot \phi(z)\right) \\
    \Pi_E &= N\left(\left(\mu-c_X\right)\cdot \Phi(-z)+\left(\frac{\sigma^2}{\nu}\right)\cdot \phi(z)\right)
\end{align}

\noindent where $\nu = \sqrt{\sigma^2+\tau^2(N)}$ and $z = (\kappa - \mu) / \nu$. Define the gap between the objectives as:

\vspace{-0.75pc}
\begin{align}
    G & \equiv   \Pi_P-\Pi_E \\
    &= N \left(c_X \cdot \Phi(-z)+ \left(\nu - \frac{\sigma^2}{\nu}\right)\cdot \phi(z)\right) \\
    &= N \left(c_X \cdot \Phi(-z)+ \left(\frac{\tau^2(N)}{\nu}\right)\cdot \phi(z)\right)
\end{align}

Assuming interior solutions, the resource constraint $R(N, \sigma^2) = \bar{R}$ binds, $N$ is determined by $\sigma^2$, and both objectives $\Pi_P$ and $\Pi_E$ may be viewed as functions of $\sigma^2$ alone. The product team's first-order condition requires that $d \Pi_P / d \sigma^2 = 0$. Since $\Pi_E=\Pi_P-G$, it follows that

\begin{equation}
    \frac{d\Pi_E}{d\sigma^2} = -\frac{dG}{d\sigma^2}
\end{equation}

\noindent Therefore, to show that the experience owner prefers higher $\sigma^2$ at the product team optimum, it suffices to show that $d G / d\sigma^2 < 0$, which implies that $d\Pi_E / d\sigma^2 > 0$. By the chain rule, we have that:

\begin{equation}
\frac{dG}{d\sigma^2} = \frac{\partial G}{\partial \sigma^2} + \frac{\partial G}{\partial N} \cdot \frac{dN}{d\sigma^2} \label{eq:prop_1_1}
\end{equation}

\noindent Since the resource constraint determines $N$ as a function of $\sigma^2$, the product team's first-order condition $d\Pi_P / d\sigma^2 = 0$ implies:

\begin{equation}
\frac{dN}{d\sigma^2} = -\frac{\partial \Pi_P/\partial \sigma^2}{\partial \Pi_P/\partial N} \label{eq:prop_1_2}
\end{equation}

Plugging equation (\ref{eq:prop_1_2}) into equation (\ref{eq:prop_1_1}) and rearranging, we have that: 

\begin{equation}
\frac{dG}{d\sigma^2} = \frac{1}{\partial \Pi_P/\partial N} \left( \frac{\partial G}{\partial \sigma^2}\frac{\partial \Pi_P}{\partial N} - \frac{\partial G}{\partial N}\frac{\partial \Pi_P}{\partial \sigma^2} \right)
\end{equation}

Because $\mu \geq 0$, $\kappa > \mu$, and $\tau^2(N)$ is strictly increasing in $N$, we have that: 

\vspace{-0.75pc} 
\begin{align}
    \frac{\partial \Pi_P}{\partial N} &= \mu \Phi\left(-z\right) + \nu \phi\left(z \right) + N \phi\left(z\right) \left(\frac{\mu z}{\nu} + 1 + z^2\right) \cdot \frac{d \tau^2(N)}{d N} \cdot \frac{1}{2 \nu} \\
    & >0 
\end{align}

It therefore suffices to show that: 

\begin{equation}
\frac{\partial G}{\partial N} \cdot \frac{\partial \Pi_P}{\partial \sigma^2} > \frac{\partial G}{\partial \sigma^2} \cdot \frac{\partial \Pi_P}{\partial N}  \label{eq:prop_1_3}
\end{equation}

To this end, note that both $G$ and $\Pi_P$ depend on $\sigma^2$ only through $\nu$, so for $F\in\{\Pi_P,G\}$:

\begin{equation}
\frac{\partial F}{\partial \sigma^2} = \frac{\partial F}{\partial \nu}\cdot \frac{1}{2 \nu}
\end{equation}

Moreover, each derivative with respect to $N$ has both a direct effect and an indirect effect through $\nu$:

\begin{equation}
\frac{\partial F}{\partial N} = \frac{\partial F}{\partial N}\Bigg|_\nu + \frac{\partial F}{\partial \nu}\cdot \frac{d \tau^2(N)}{dN} \cdot \frac{1}{2 \nu}
\end{equation}

Applying both of these facts to equation (\ref{eq:prop_1_3}) and simplifying, we see that it therefore suffices to show that: 

\begin{equation}
\frac{\partial G}{\partial N}\Bigg|_\nu \cdot \frac{\partial \Pi_P}{\partial \nu} >  \frac{\partial G}{\partial \nu} \cdot \frac{\partial \Pi_P}{\partial N}\Bigg|_\nu  \label{eq:prop_1_4}
\end{equation}

Evaluating each derivative, we have that: 

\vspace{-0.75pc}
\begin{align}
   \frac{\partial G}{\partial N}\Bigg|_\nu  &= c_X \cdot \Phi(-z)+\left(\frac{\tau^2(N)+N \cdot d \tau^{2}(N)/dN}{\nu}\right)\phi(z) \\
   \frac{\partial \Pi_P}{\partial \nu} &= N\cdot \phi(z) \cdot \left(\frac{\mu z}{\nu}+1+z^2\right) \\
    \frac{\partial G}{\partial \nu} &= N \cdot \phi\left(z\right) \cdot  \left(\frac{c_X z}{\nu} +\frac{\tau^2(N) \cdot (z^2-1) }{\nu^2}\right)\\
    \frac{\partial \Pi_P}{\partial N}\Bigg|_\nu &= \mu \cdot \Phi(-z) + \nu \cdot \phi(z)
\end{align}

Substituting these expressions into equation (\ref{eq:prop_1_4}), we have that the difference between the left and right-hand sides is given by: 

\vspace{-0.75pc}
\begin{align}
    \frac{\partial G}{\partial N}\Bigg|_\nu \cdot \frac{\partial \Pi_P}{\partial \nu} -  \frac{\partial G}{\partial \nu} \cdot \frac{\partial \Pi_P}{\partial N}\Bigg|_\nu = N\phi\left(z\right) \Bigg(&c_X\left(\left(1+z^2\right)\Phi\left(-z\right)-z\phi\left(z\right)\right) \label{eq:prop_1_5} \\
    &+ \left(\frac{N \cdot d \tau^{2}(N) / dN}{\nu}\right)\phi\left(z\right)\left(1+z^2\right) \label{eq:prop_1_6}\\
    &+ \left(\frac{N \mu  z \cdot d \tau^{2}(N) / dN}{\nu^2}\right)\phi(z) + \left(\frac{2\tau^2(N)}{\nu}\right) \phi(z) \label{eq:prop_1_7}\\
    &+ \left(\frac{\tau^2(N) \cdot \mu}{\nu^2}\right)\left[z\cdot \phi(z)-\Phi(-z)(z^2-1)\right]\Bigg) \label{eq:prop_1_8}
\end{align}

This difference is positive. That the term in equation (\ref{eq:prop_1_5}) is non-negative follows from the assumption that $c_X \geq 0$ and the Mills' ratio bound $\Phi(-z) \geq z(1+z^2)^{-1} \cdot \phi(z)$ for $z>0$ \citep{gordon_1941}. That the terms in equations (\ref{eq:prop_1_6}) and (\ref{eq:prop_1_7}) are positive follows from the assumptions that $\mu \geq 0$, $\kappa > \mu$, and $d \tau^2(N)/dN > 0$. For the term in equation (\ref{eq:prop_1_8}), note that if $0<z \leq 1$, then $z^2-1 \leq 0$, so the bracketed term is positive; if $z>1$, the Mills' ratio bound $\Phi(-z) \le \phi(z)/z$ implies $\Phi(-z)(z^2-1) < z\phi(z)$, so the bracketed term is again positive \citep{gordon_1941}. This together with the assumption that $\mu \geq 0$ guarantees that the term in equation (\ref{eq:prop_1_8}) is non-negative. 

\noindent It follows that $d\Pi_E / d\sigma^2 >0$ at the product team optimum. Since $N$ is decreasing in $\sigma^2$ along the binding resource frontier, this is equivalent to the experience owner's objective being decreasing in $N$ at that point. Because $\tau^2(N)$ is strictly increasing in $N$, this marginal reduction in $N$ also raises statistical precision. Therefore, the product team's raw-scoring allocation can be improved for the experience owner by a marginal reallocation toward fewer, more precise tests of more ambitious ideas $\hfill \square$

\vspace{0.5pc}

\noindent \textbf{Proposition 2 (Vanishing testing costs under raw scoring):} Suppose that $\mu \geq 0$, $\kappa > \mu$, and the resource function takes the form $R(N,\sigma^2)=\gamma N+R_\sigma(\sigma^2)$, where $\gamma>0$. Assume that $R_\sigma$ is continuous and strictly increasing, with $R_\sigma(0)=0$ and $R_\sigma(\bar x)=\bar R$ for some finite $\bar x>0$. Suppose also that $\tau^2(N)\to\infty$ as $N\to\infty$. Under equations (\ref{eq:prior}) and (\ref{eq:sampling}) and raw scoring $s(\hat{\Delta},\tau^2,\sigma^2)=\hat{\Delta}$, any optimal product-team allocation satisfies $\tilde{\sigma}^2\to0$ and $\gamma\tilde N\to\bar R$ as $\gamma\to0$. Thus, the team approaches minimum ambition and maximum feasible test volume, while sampling variance diverges and statistical precision vanishes.

\noindent \emph{Proof:} Write $x=\sigma^2$. Along the binding resource frontier, $N=(\bar R-R_\sigma(x))/\gamma$. Since $\partial\Pi_P/\partial N>0$ (shown in Proposition 1), the product team’s resource constraint binds. Define $N_{\max}=\bar R/\gamma$, the maximum feasible test volume, attained at $x=0$.

\noindent Under raw scoring, write the product team’s objective as
\begin{equation}
\Pi_P(N,x)=N f\!\left(\sqrt{x+\tau^2(N)}\right),
\end{equation}
\noindent where
\begin{equation}
f(v)=\mu\Phi\!\left(-\frac{\kappa-\mu}{v}\right)
     +v\phi\!\left(\frac{\kappa-\mu}{v}\right).
\end{equation}
For $\mu\geq0$ and $\kappa>\mu$, $f(v)$ is strictly increasing for $v>0$ (this is the same derivative $\partial \Pi_P^R/\partial \nu = N\phi(z_R) A_R > 0$ established in the proof of Proposition 4, with $v=\nu$), and
\begin{equation}
\frac{f(v)}{v}\longrightarrow\phi(0)
\qquad\text{as }v\to\infty.
\end{equation}

Fix any $\epsilon\in(0,\bar x)$. Every feasible allocation with $x\geq\epsilon$ satisfies $x\leq\bar x$ and
\begin{equation}
N\leq c_\epsilon N_{\max},
\qquad
c_\epsilon=1-\frac{R_\sigma(\epsilon)}{\bar R}\in(0,1).
\end{equation}
Because $\tau^2(N)$ and $f(v)$ are increasing, its payoff relative to the feasible allocation $(N_{\max},0)$ satisfies
\begin{equation}
\frac{\Pi_P(N,x)}{\Pi_P(N_{\max},0)}
\leq
c_\epsilon
\frac{f\!\left(\sqrt{\bar x+\tau^2(N_{\max})}\right)}
     {f\!\left(\sqrt{\tau^2(N_{\max})}\right)}.
\end{equation}
As $\gamma\to0$, $N_{\max}\to\infty$ and $\tau^2(N_{\max})\to\infty$. The ratio of the two arguments of $f$ tends to one, so the displayed upper bound converges to $c_\epsilon<1$. Thus, for sufficiently small $\gamma$, no optimal allocation can have $x\geq\epsilon$. Since this holds for every $\epsilon\in(0,\bar x)$, $\tilde{\sigma}^2\to0$.

The binding resource constraint and continuity of $R_\sigma$ then imply
\begin{equation}
\gamma\tilde N
=\bar R-R_\sigma(\tilde{\sigma}^2)
\longrightarrow\bar R.
\end{equation}
Consequently, $\tilde N/N_{\max}\to1$ and $\tilde N\to\infty$. Since $\tau^2(N)\to\infty$ as $N\to\infty$, sampling variance diverges and statistical precision vanishes $\hfill \square$

\vspace{0.5pc}

\noindent \textbf{Proposition 3 (Posterior-mean scoring removes the reward to noise):} Suppose that $\mu \geq 0$ and $\kappa > \mu$. Under equations (\ref{eq:prior}) and (\ref{eq:sampling}) and posterior-mean scoring $s(\hat{\Delta}, \tau^2, \sigma^2) = \mu + \sigma^2(\sigma^2 + \tau^2)^{-1}(\hat{\Delta} - \mu)$, the gap $G = \Pi_P - \Pi_E$ between the product team's and the experience owner's objectives is strictly smaller for any $(N, \sigma^2)$ than under raw scoring. When $c_X = 0$, $G = 0$ for all $(N, \sigma^2)$.

\noindent \emph{Proof:} Fix the launch threshold $\kappa$ and assume that $\mu \geq 0$ and $\kappa  > \mu$. If $\sigma^2=0$, posterior-mean scoring gives the constant score $\mu<\kappa$, so no features launch and $G_M=0$. Under raw scoring, for $N>0$ and $\tau^2(N)>0$,
\begin{equation}
G_R=N\left[
c_X\Phi\!\left(\frac{\mu-\kappa}{\tau(N)}\right)
+\tau(N)\phi\!\left(\frac{\kappa-\mu}{\tau(N)}\right)
\right]>0.
\end{equation}
Thus, the claimed comparison also holds at $\sigma^2=0$. For the remainder of the proof, suppose that $\sigma^2>0$. By Lemmas 1 and 2, we have that for any positive-weight linear shrinkage scoring:

\vspace{-0.75pc}
{\footnotesize
\begin{align}
G &\equiv \Pi_P - \Pi_E \\
&= N\left( c_X \cdot \Phi\!\left(\frac{\mu-\kappa}{w(N,\sigma^2)\nu(N,\sigma^2)}\right) + \left( w(N,\sigma^2) \nu(N,\sigma^2)-\frac{\sigma^2}{\nu(N,\sigma^2)} \right) \cdot \phi\!\left(\frac{\kappa-\mu}{w(N,\sigma^2)\nu(N,\sigma^2)}\right) \right)
\end{align}}

\noindent Under raw scoring $s(\hat{\Delta}, \tau^2, \sigma^2) = \hat{\Delta}$, this simplifies to:

\begin{equation}
G_R = N\left( c_X \cdot \Phi\!\left(\frac{\mu-\kappa}{\nu(N,\sigma^2)}\right) + \left(\frac{\tau^2(N)}{\nu(N,\sigma^2)}\right) \cdot \phi\!\left(\frac{\kappa-\mu}{\nu(N,\sigma^2)}\right) \right)
\end{equation}

while under posterior-mean scoring $s(\hat{\Delta}, \tau^2, \sigma^2) = \mu + \sigma^2(\sigma^2 + \tau^2)^{-1}(\hat{\Delta} - \mu)$ it simplifies to: 

\begin{equation}
G_M = N c_X \cdot \Phi\!\left(\frac{\mu-\kappa}{w(N,\sigma^2)\nu(N,\sigma^2)}\right)
\end{equation}

\noindent with $w(N, \sigma^2) = w(\tau^2(N), \sigma^2) = \sigma^2 (\sigma^2 + \tau^2(N))^{-1}$. Since $\tau^2(N) > 0$, we have $w(N,\sigma^2) < 1$. Under the assumption that $\kappa > \mu$, this implies:

\begin{equation}
\frac{\mu-\kappa}{w(N,\sigma^2)\nu(N,\sigma^2)} <  \frac{\mu-\kappa}{\nu(N,\sigma^2)}
\end{equation} 

and therefore that: 

\begin{equation}
    G_M \leq N c_X \cdot \Phi\!\left(\frac{\mu-\kappa}{\nu(N,\sigma^2)}\right) < G_R
\end{equation}

\noindent which establishes the claim. Finally, note that if $c_X=0$, then $G_M = 0$ and $\Pi_P = \Pi_E$ for all $(N, \sigma^2)$ $\hfill \square$

\vspace{0.5pc}

\noindent \textbf{Proposition 4 (Scoring rules change marginal allocation incentives):} Suppose that $\mu \geq 0$ and $\kappa > \mu$. Under equations (\ref{eq:prior}) and (\ref{eq:sampling}), at an interior product-team optimum under raw scoring $s(\hat{\Delta}, \tau^2, \sigma^2) = \hat{\Delta}$, the product team's objective under posterior-mean scoring $s(\hat{\Delta}, \tau^2, \sigma^2) = \mu + \sigma^2(\sigma^2 + \tau^2)^{-1}(\hat{\Delta} - \mu)$ has a strictly positive derivative with respect to $\sigma^2$ along the binding resource frontier. Thus, under posterior-mean scoring, the team prefers a marginal reallocation toward fewer, more precise tests of more ambitious ideas.

\noindent \emph{Proof:} Fix the launch threshold $\kappa$ and assume that $\mu \geq 0$ and $\kappa > \mu$. By Lemma 1, we have that under raw and posterior-mean scoring:

\vspace{-0.75pc}
\begin{align}
    \Pi_P^R &= N\left(\mu \cdot \Phi(-z_R) + \nu \cdot \phi(z_R)\right) \\
    \Pi_P^M &= N\left(\mu \cdot \Phi(-z_M) + \left(\frac{\sigma^2}{\nu}\right) \cdot \phi(z_M)\right)
\end{align}

\noindent where $\nu = \sqrt{\sigma^2+\tau^2(N)}$, $z_R = (\kappa - \mu)/\nu$, $z_M = (\kappa - \mu)\nu/\sigma^2 = z_R / w$, and $w = \sigma^2/\nu^2 \in (0,1)$.

Assuming interior solutions, the resource constraint $R(N, \sigma^2) = \bar{R}$ binds, $N$ is determined by $\sigma^2$, and both objectives may be viewed as functions of $\sigma^2$ alone. At the raw-scoring optimum, $d\Pi_P^R / d\sigma^2 = 0$. To establish this local reallocation incentive, it suffices to show that $d\Pi_P^M / d\sigma^2 > 0$ at the raw-scoring optimum. By the chain rule, we have that:

\begin{equation}
\frac{d\Pi_P^M}{d\sigma^2} = \frac{\partial \Pi_P^M}{\partial \sigma^2} + \frac{\partial \Pi_P^M}{\partial N} \cdot \frac{dN}{d\sigma^2} \label{eq:prop_4_1}
\end{equation}

\noindent Since the resource constraint determines $N$ as a function of $\sigma^2$, the product team's first-order condition $d\Pi_P^R / d\sigma^2 = 0$ implies:

\begin{equation}
\frac{dN}{d\sigma^2} = -\frac{\partial \Pi_P^R/\partial \sigma^2}{\partial \Pi_P^R/\partial N} \label{eq:prop_4_2}
\end{equation}

Plugging equation (\ref{eq:prop_4_2}) into equation (\ref{eq:prop_4_1}) and rearranging, we have that:

\begin{equation}
\frac{d\Pi_P^M}{d\sigma^2} = \frac{1}{\partial \Pi_P^R/\partial N} \left( \frac{\partial \Pi_P^M}{\partial \sigma^2}\frac{\partial \Pi_P^R}{\partial N} - \frac{\partial \Pi_P^M}{\partial N}\frac{\partial \Pi_P^R}{\partial \sigma^2} \right)
\end{equation}

We showed in the proof of Proposition 1 that $\partial \Pi_P^R / \partial N > 0$. It therefore suffices to show that:

\begin{equation}
\frac{\partial \Pi_P^M}{\partial \sigma^2} \cdot \frac{\partial \Pi_P^R}{\partial N} > \frac{\partial \Pi_P^M}{\partial N} \cdot \frac{\partial \Pi_P^R}{\partial \sigma^2}  \label{eq:prop_4_3}
\end{equation}

Note that $\Pi_P^R$ depends on $\sigma^2$ and $\tau^2$ only through $\nu$, while $\Pi_P^M$ depends on both $\nu$ and $\sigma^2$ directly (through $w$). For either objective $F \in \{\Pi_P^R, \Pi_P^M\}$, the derivative with respect to $N$ has both a direct effect and an indirect effect through $\nu$:

\begin{equation}
\frac{\partial F}{\partial N} = \frac{\partial F}{\partial N}\Bigg|_\nu + \frac{\partial F}{\partial \nu}\cdot \frac{d \tau^2(N)}{dN} \cdot \frac{1}{2 \nu}
\end{equation}

Similarly, the derivative of $\Pi_P^R$ with respect to $\sigma^2$ operates entirely through $\nu$:

\begin{equation}
\frac{\partial \Pi_P^R}{\partial \sigma^2} = \frac{\partial \Pi_P^R}{\partial \nu}\cdot \frac{1}{2 \nu}
\end{equation}

while the derivative of $\Pi_P^M$ with respect to $\sigma^2$ has both a direct effect (through $w$) and an indirect effect (through $\nu$):

\begin{equation}
\frac{\partial \Pi_P^M}{\partial \sigma^2} = \frac{\partial \Pi_P^M}{\partial \sigma^2}\Bigg|_\nu + \frac{\partial \Pi_P^M}{\partial \nu}\cdot \frac{1}{2 \nu}
\end{equation}

Evaluating these derivatives, we have that:

\vspace{-0.75pc}
\begin{align}
   \frac{\partial \Pi_P^R}{\partial N}\Bigg|_\nu &= \mu \cdot \Phi(-z_R) + \nu \cdot \phi(z_R) \\
   \frac{\partial \Pi_P^R}{\partial \nu} &= N\cdot \phi(z_R) \cdot A_R \\
   \frac{\partial \Pi_P^M}{\partial N}\Bigg|_\nu &= \mu \cdot \Phi(-z_M) + \frac{\sigma^2}{\nu} \cdot \phi(z_M) \\
   \frac{\partial \Pi_P^M}{\partial \sigma^2}\Bigg|_\nu &= \left(\frac{N \phi(z_M)}{\nu w}\right) \cdot A_M \\
   \frac{\partial \Pi_P^M}{\partial \nu} &= -N \cdot \phi(z_M) \cdot A_M
\end{align}

\noindent where $A_R = \mu z_R / \nu + 1 + z_R^2 > 0$ and $A_M = \mu z_M / \nu + w(1 + z_M^2) > 0$.

Substituting the decompositions into equation (\ref{eq:prop_4_3}), expanding, and noting that the terms involving $(\partial \Pi_P^M / \partial \nu) \cdot (\partial \Pi_P^R / \partial \nu)$ cancel between the two sides, we see that it suffices to show that: 

\begin{equation}
\Theta = \frac{\partial \Pi_P^M}{\partial \sigma^2}\Bigg|_\nu \cdot \frac{\partial \Pi_P^R}{\partial N} + \frac{1}{2\nu}\left(\frac{\partial \Pi_P^M}{\partial \nu} \cdot \frac{\partial \Pi_P^R}{\partial N}\Bigg|_\nu - \frac{\partial \Pi_P^M}{\partial N}\Bigg|_\nu \cdot \frac{\partial \Pi_P^R}{\partial \nu}\right) > 0 \label{eq:prop_4_reduced}
\end{equation}

Plugging in the derivatives we derived above, we have that:

\vspace{-0.75pc}
\begin{align}
    \Theta = \frac{N}{2w\nu}\Bigg(&\phi(z_R) \cdot \phi(z_M) \cdot \nu \left((2-w) A_M - w^2 A_R\right) \label{eq:prop_4_t1}\\
    &+ \phi(z_R) \cdot \phi(z_M) \left( \frac{N \cdot d\tau^2(N)/dN}{\nu} \right) \cdot A_R A_M \label{eq:prop_4_t2}\\
    &+ \mu \left(\phi(z_M)(2-w) A_M \cdot \Phi(-z_R) - w \cdot \phi(z_R) A_R \cdot \Phi(-z_M)\right) \Bigg) \label{eq:prop_4_t3}
\end{align}

\noindent Since $N/(2w\nu) > 0$, it suffices to show that the term in parentheses is positive. That the term in equation (\ref{eq:prop_4_t1}) is strictly positive follows from the identity:

\vspace{-0.75pc}
\begin{equation}
    (2-w) A_M - w^2 A_R = \left(\frac{\mu z_R}{\nu} + z_R^2\right) \cdot \left(\frac{2 - w - w^3}{w}\right) + 2w(1-w)
\end{equation}

\noindent where $2 - w - w^3 > 0$ for $w \in (0,1)$ (the function equals $2$ at $w=0$, equals $0$ at $w=1$, and has strictly negative derivative $-1-3w^2$). That the term in equation (\ref{eq:prop_4_t2}) is positive follows from the assumptions that $d\tau^2(N)/dN > 0$ and $A_R, A_M > 0$. That the term in equation (\ref{eq:prop_4_t3}) is non-negative follows from two facts: first, the inverse Mills ratio $\phi(z)/\Phi(-z)$ is strictly increasing in $z$ for $z > 0$ \citep{gordon_1941}, and since $z_M = z_R / w > z_R$ we have $\phi(z_M) \cdot \Phi(-z_R) > \phi(z_R) \cdot \Phi(-z_M)$; second, $(2-w) A_M > w A_R$, which follows from the identity $(2-w) A_M - w A_R = \frac{(1-w)(2+w)}{w}\left(\frac{\mu z_R}{\nu} + z_R^2\right) + w(1-w) > 0$ for $w \in (0,1)$.

\noindent It follows that $d\Pi_P^M/d\sigma^2>0$ at the raw-scoring optimum. A marginal increase in $\sigma^2$ along the binding resource frontier therefore raises the product team's posterior-mean-scoring objective while reducing test volume and improving statistical precision. $\hfill \square$

\clearpage 

\section{Appendix: Supplemental Results}

\begin{figure}[h!]
    \caption{Sensitivity of empirical findings across samples and outcomes}
    \label{fig:robustness}
    \centering 
    \begin{tabular}{cccc}
       {\fontsize{4}{5}\selectfont (a) Metric A (Excluding extreme $\hat{\Delta}$)} &  {\fontsize{4}{5}\selectfont (b) Metric A (Excluding extreme $\tau^2$)}  & {\fontsize{4}{5}\selectfont (c) Metric A (Excluding extreme $ \hat{\Delta}/\tau$) }  & {\fontsize{4}{5}\selectfont (d) Metric A (Excluding largest product teams)} \\
       \includegraphics[scale = 0.22]{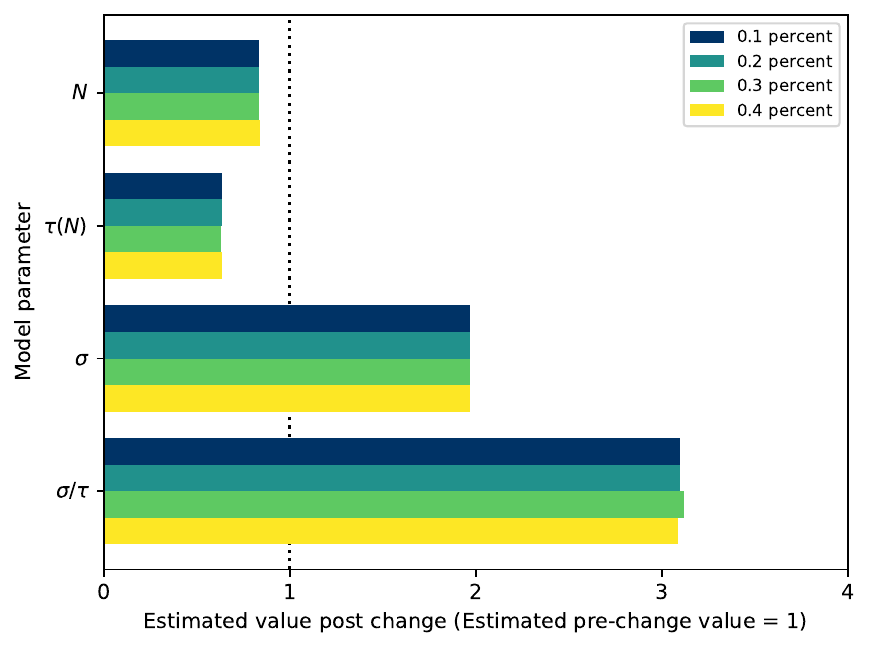} & \includegraphics[scale = 0.22]{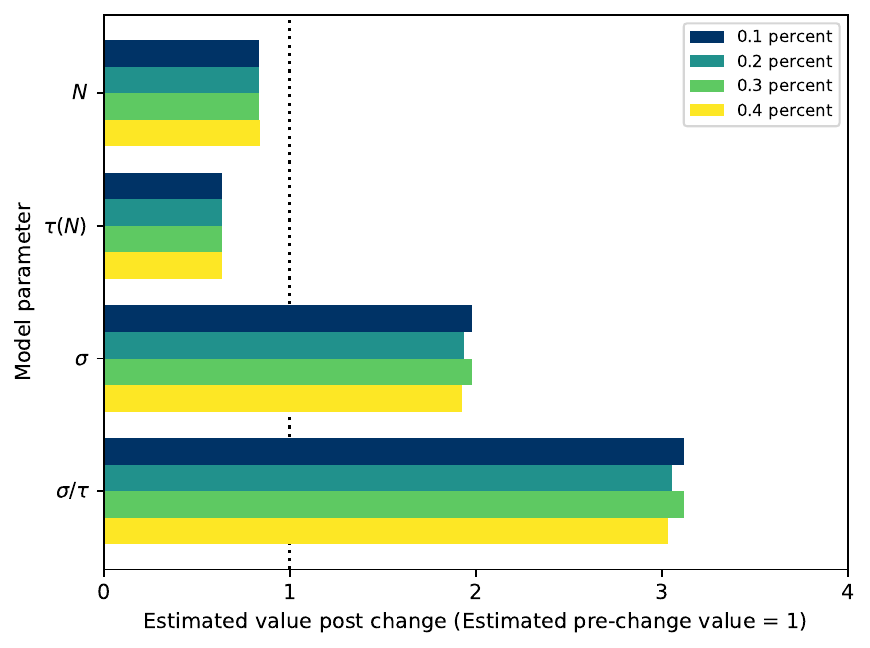} &  \includegraphics[scale = 0.22]{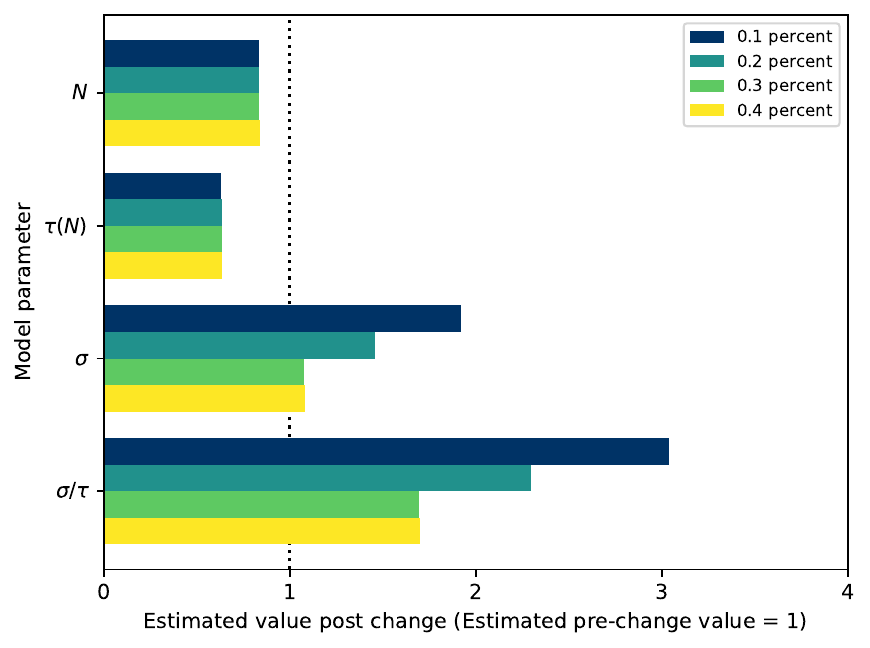}  & \includegraphics[scale = 0.22]{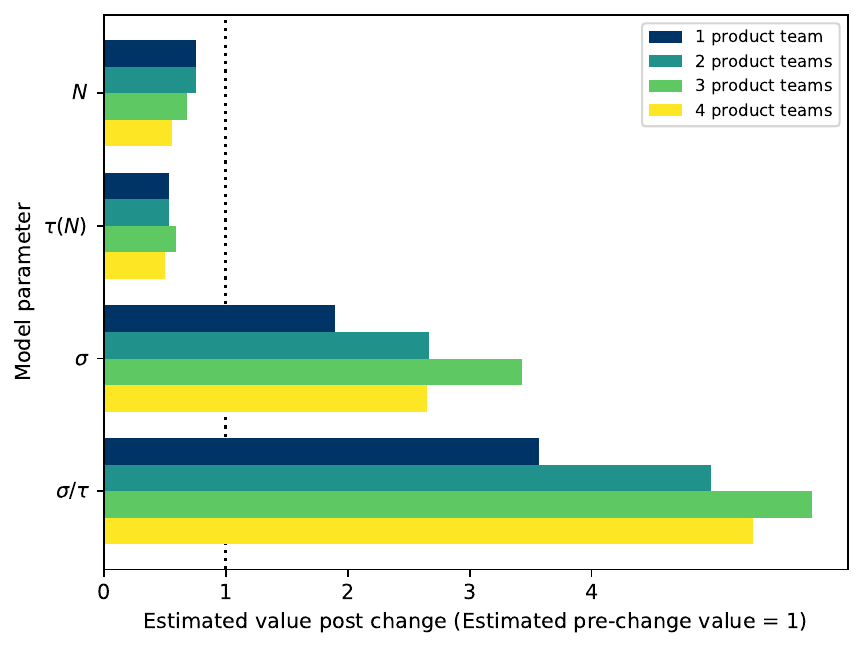}  \\
       {\fontsize{4}{5}\selectfont (e) Metric A (Product team reweighting)} & {\fontsize{4}{5}\selectfont (f) Metric A (Expanded time period)} &  {\fontsize{4}{5}\selectfont (g) Metric B (Excluding extreme $\hat{\Delta}$)}  & {\fontsize{4}{5}\selectfont (h) Metric B (Excluding extreme $\tau^2$)}  \\
       \includegraphics[scale = 0.22]{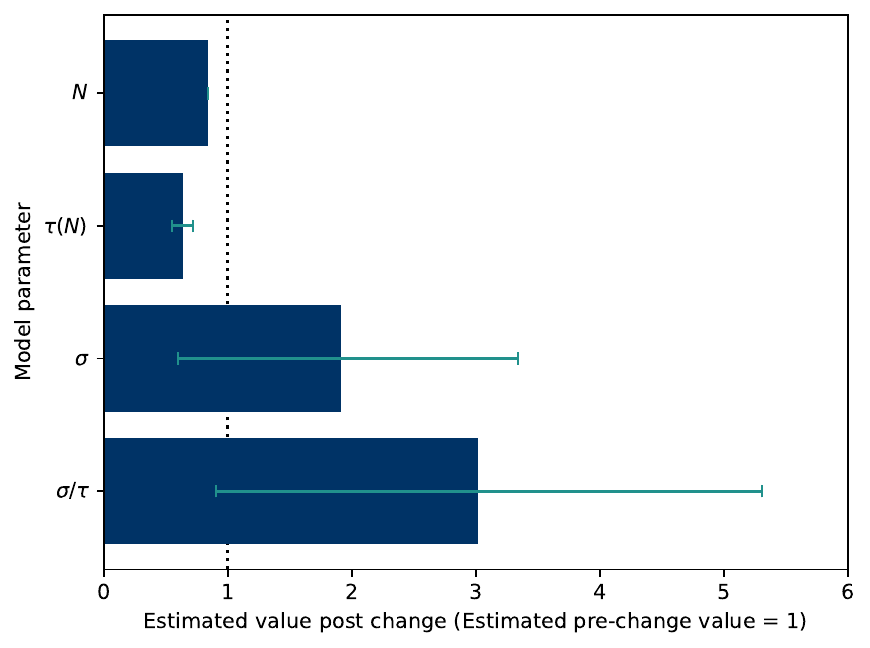} & \includegraphics[scale = 0.22]{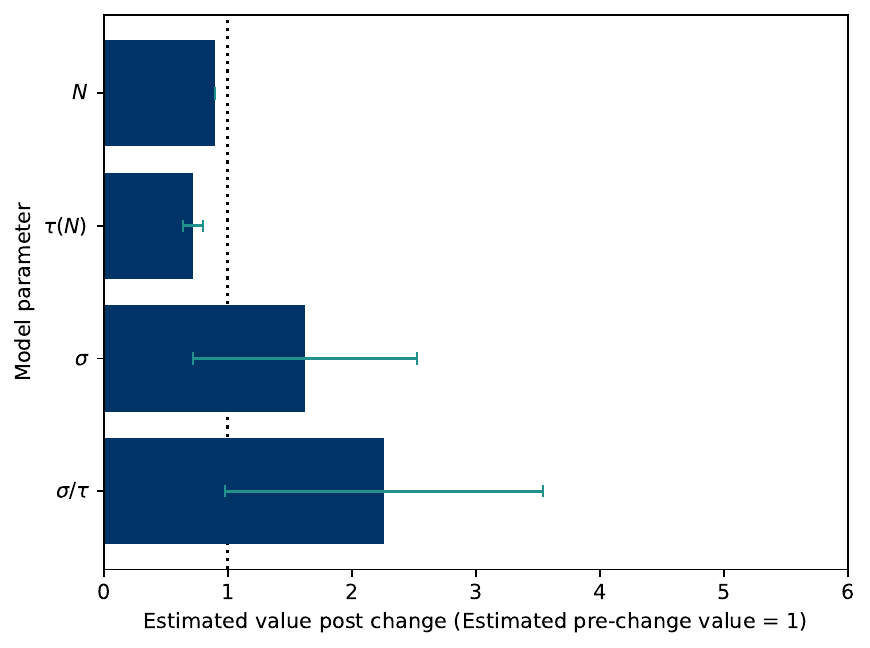} &  \includegraphics[scale = 0.22]{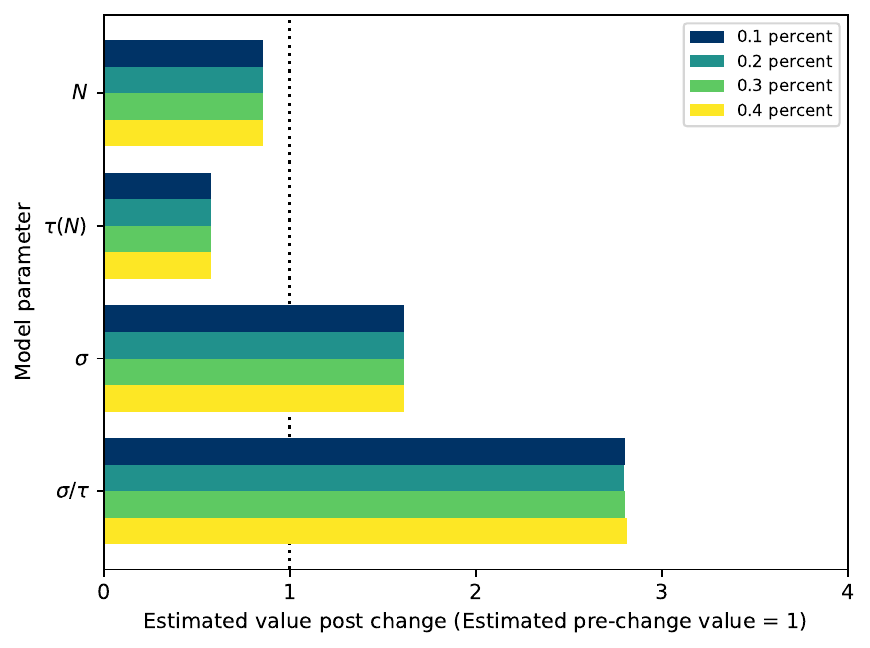} & \includegraphics[scale = 0.22]{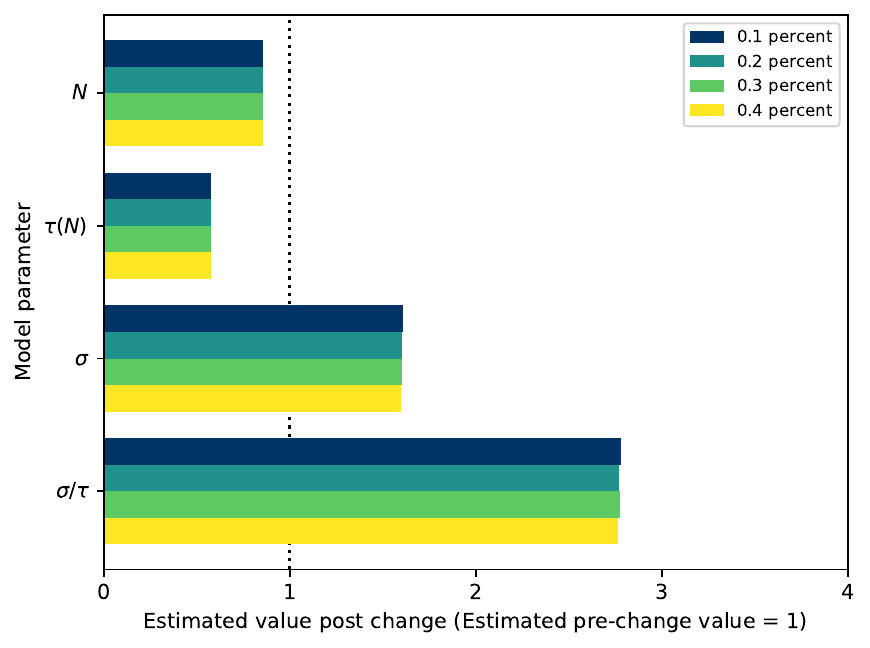} \\
        {\fontsize{4}{5}\selectfont (i) Metric B (Excluding extreme $ \hat{\Delta}/\tau$)}  & {\fontsize{4}{5}\selectfont (j) Metric B (Excluding largest product teams)} & {\fontsize{4}{5}\selectfont (k) Metric B (Product team reweighting)} & {\fontsize{4}{5}\selectfont (l) Metric B (Expanded time period)} \\
        \includegraphics[scale = 0.22]{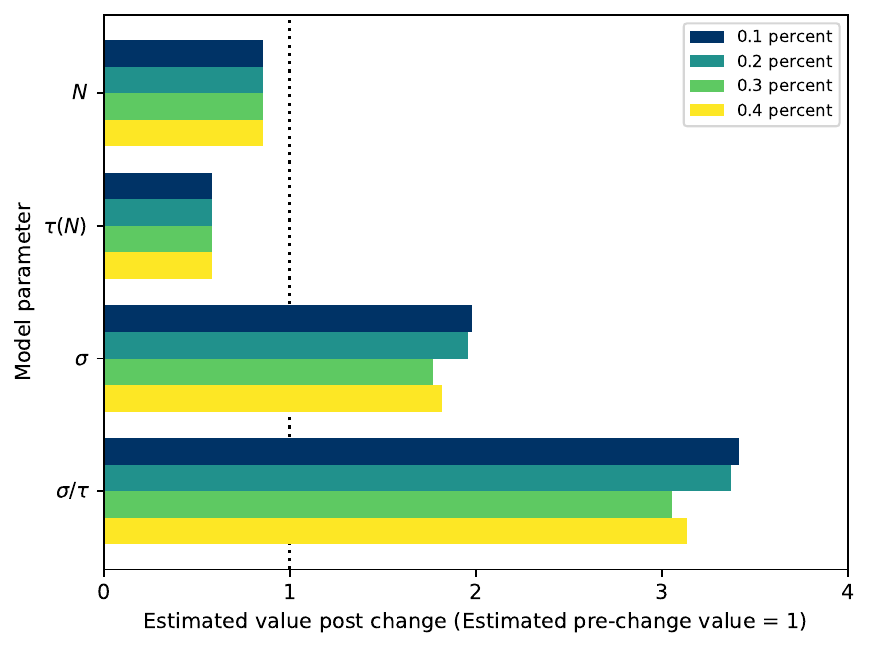} & \includegraphics[scale = 0.22]{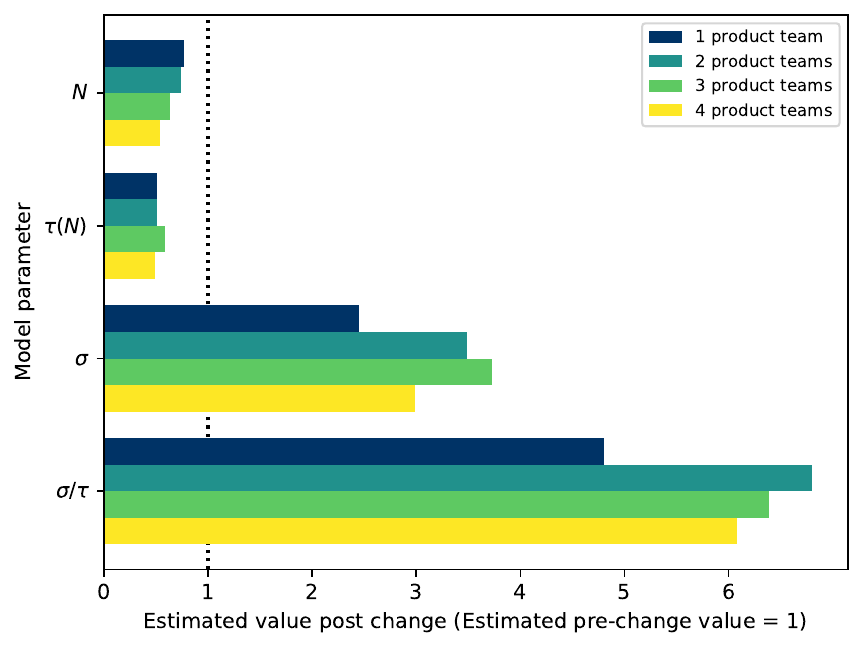} & \includegraphics[scale = 0.22]{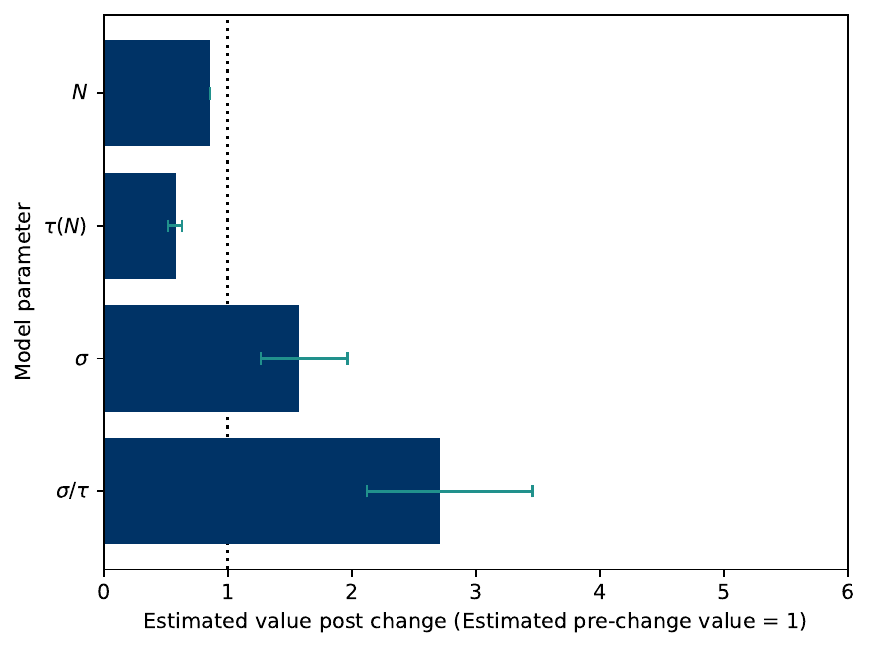} & \includegraphics[scale = 0.22]{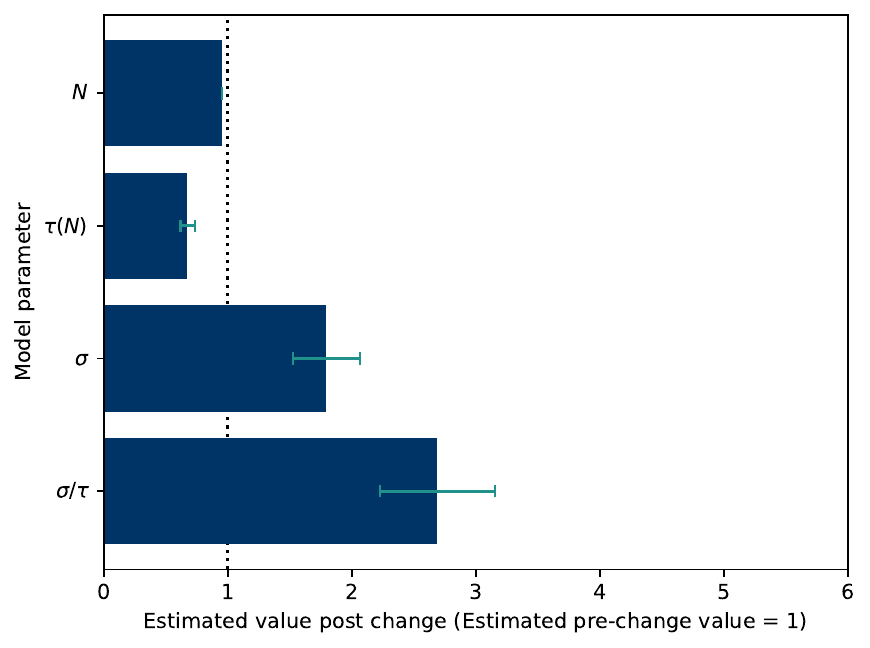} \\
         {\fontsize{4}{5}\selectfont (m) Metric C} &  {\fontsize{4}{5}\selectfont (n) Metric D} &  {\fontsize{4}{5}\selectfont (o) Metric E} &  {\fontsize{4}{5}\selectfont (p) Metric F (Falsification)}  \\
          \includegraphics[scale = 0.22]{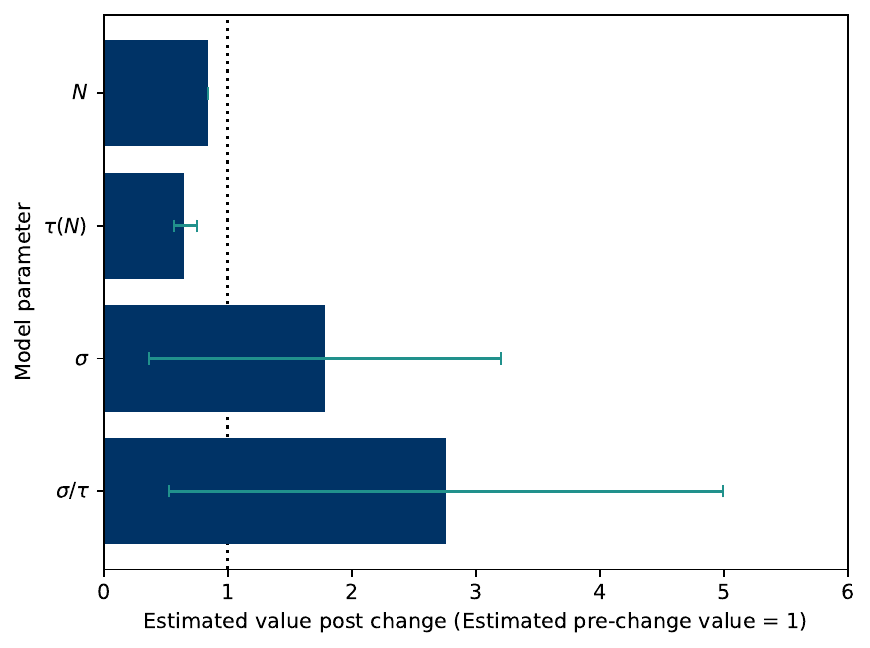} &  \includegraphics[scale = 0.22]{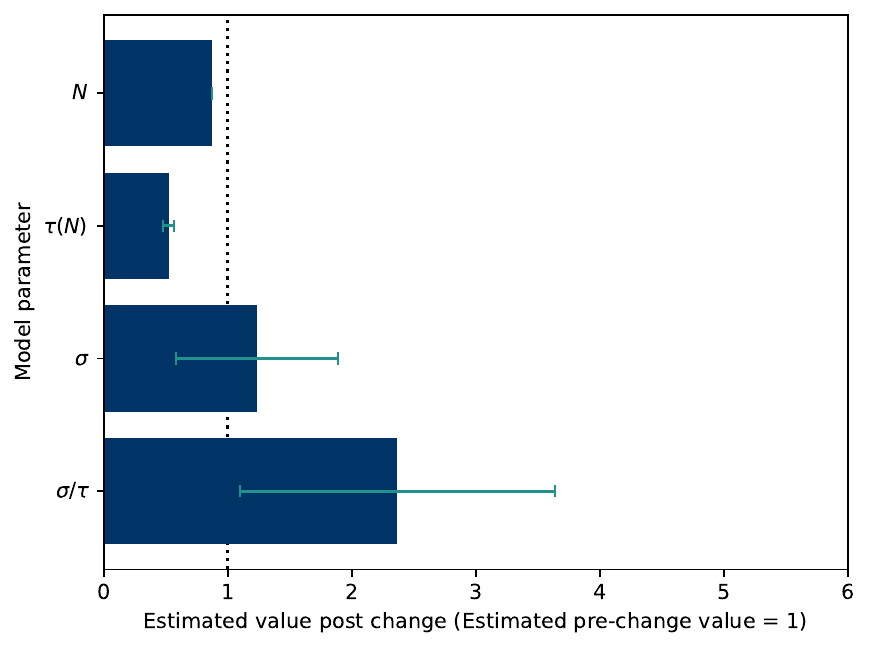} &  \includegraphics[scale = 0.22]{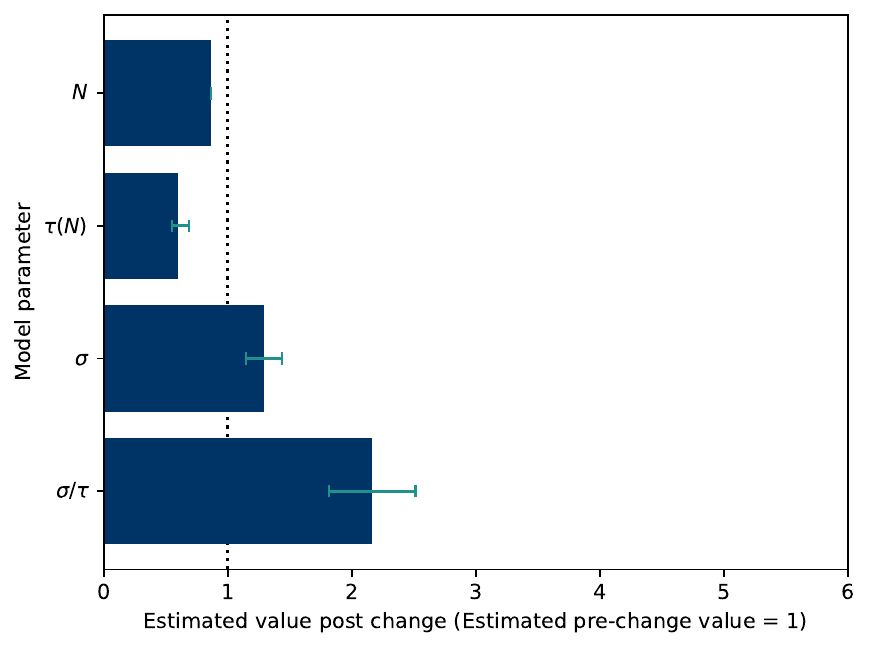} &  \includegraphics[scale = 0.22]{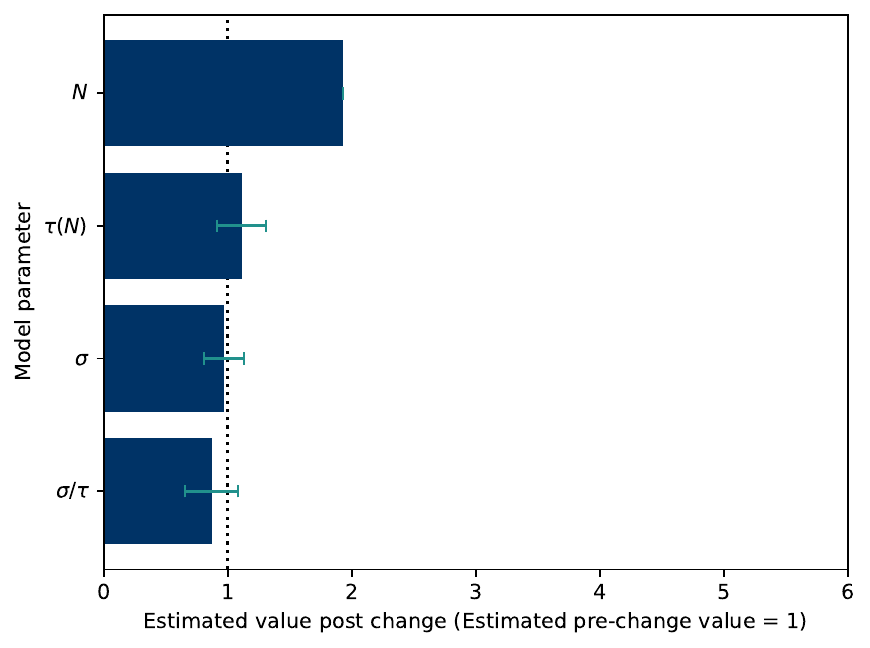}
    \end{tabular}
    \begin{spacing}{0.9}
    \begin{flushleft}
        \tiny
        \emph{Notes:} In panel (a), the outcome metric is metric A but test-treatments with estimated impacts in the top and bottom 0.1, 0.2, 0.3, and 0.4 percent are excluded from both the pre- and post-change periods. In panel (b), the outcome metric is metric A but test-treatments with sampling variances in the top and bottom 0.1, 0.2, 0.3, and 0.4 percent are excluded from both the pre- and post-change periods. In panel (c), the outcome metric is metric A but test-treatments with estimated $t$-statistics in the top and bottom 0.1, 0.2, 0.3, and 0.4 percent are excluded from both the pre- and post-change periods. In panel (d), the outcome metric is metric A but the largest 1, 2, 3, and 4 product teams are excluded from both the pre- and post-change periods. In panel (e), the outcome metric is metric A but the sample is re-weighted such that the distribution of test-treatments across product teams in the post-change period matches that of the pre-change period. In panel (f), the outcome metric is metric A but the pre-change period is expanded to 2023-08-01 to 2024-02-28 (instead of 2023-09-01 to 2024-01-31) and the post-change period is expanded to 2024-08-01 to 2025-02-28 (instead of 2024-09-01 to 2025-01-31). Panels (g)--(l) repeat the robustness checks in panels (a)--(f) using metric B as the outcome metric. In panel (m), the outcome metric is metric C. In panel (n), the outcome metric is metric D. In panel (o), the outcome metric is metric E. In panel (p), the outcome metric is metric F and the sample is test-treatment pairs run on a different Amazon experience that continued to use raw scoring throughout 2023--2024.
    \end{flushleft}
    \end{spacing}
\end{figure}

\end{document}